\documentclass[oneside,twocolumn,9.5pt]{article}
\usepackage[margin=1.0in]{geometry}
\usepackage{mathptmx}
\usepackage{graphicx}
\usepackage[semicolon]{natbib}
\usepackage{aas_macros_arxiv_hacked}
\usepackage{amssymb}
\usepackage{amsmath}
\usepackage[colorlinks,bookmarks=false,linkcolor=blue,urlcolor=red,citecolor=blue]{hyperref}
\usepackage{afterpage}

\newcommand*\mean[1]{\overline{#1}}

\title{Searching for biosignatures in exoplanetary impact ejecta}
\author{Gianni Cataldi$^{1,2}$ \and Alexis Brandeker$^{1,2}$ \and Philippe Th\'ebault$^3$ \and Kelsi Singer$^4$ \and Engy Ahmed$^{5,6,2}$ \and Bernard L. de Vries$^{1,2,7}$ \and Anna Neubeck$^{6,2}$ \and G\"oran Olofsson$^{1,2}$}
\date{}

\begin{document}

\twocolumn[
\begin{@twocolumnfalse}
\maketitle
$^1$AlbaNova University Centre, Stockholm University, Department of Astronomy, SE-106 91 Stockholm, Sweden\\
$^2$Stockholm University Astrobiology Centre, SE-106 91 Stockholm, Sweden\\
$^3$LESIA-Observatoire de Paris, UPMC Univ.\ Paris 06, Univ.\ Paris-Diderot, France\\
$^4$Southwest Research Institute, 1050 Walnut Street, Suite 300, Boulder, CO 80302, USA\\
$^5$Royal Institute of Technology (KTH), Science for Life Laboratory, SE-17121 Solna, Sweden\\
$^6$Stockholm University, Department of Geological Sciences, SE-10691 Stockholm, Sweden\\
$^7$Scientific Support Office, Directorate of Science, European Space Research and Technology Centre (ESA/ESTEC), Keplerlaan 1, 2201 AZ Noordwijk, The Netherlands\\
\begin{abstract}
\noindent
With the number of confirmed rocky exoplanets increasing steadily, their characterisation and the search for exoplanetary biospheres is becoming an increasingly urgent issue in astrobiology. To date, most efforts have concentrated on the study of exoplanetary atmospheres. Instead, we aim to investigate the possibility of characterising an exoplanet (in terms of habitability, geology, presence of life etc.) by studying material ejected from the surface during an impact event. For given parameters characterising the impact event, we estimate the escaping mass and assess its subsequent collisional evolution in a circumstellar orbit, assuming a Sun-like host star. We calculate the fractional luminosity of the dust as a function of time after the impact event and study its detectability with current and future instrumentation. We consider the possibility to constrain the dust \emph{composition}, giving information on the geology or the presence of a biosphere. As examples, we investigate whether calcite, silica or ejected microorganisms could be detected. For a 20\,km diameter impactor, we find that the dust mass escaping the exoplanet is roughly comparable to the zodiacal dust, depending on the exoplanet size. The collisional evolution is best modelled by considering two independent dust populations, a \textit{spalled} population consisting of non-melted ejecta evolving on timescales of millions of years, and dust \textit{recondensed} from melt or vapour evolving on much shorter timescales. While the presence of dust can potentially be inferred with current telescopes, studying its composition requires advanced instrumentation not yet available. The direct detection of biological matter turns out to be extremely challenging. Despite considerable difficulties (small dust masses, noise such as exozodiacal dust etc.), studying dusty material ejected from an exoplanetary surface might become an interesting complement to atmospheric studies in the future.
\end{abstract}
\end{@twocolumnfalse}
]

\section{Introduction}
Although terrestrial planets orbiting solar-type main-sequence stars seem to be relatively common in the Galaxy \citep[e.g.][]{Fressin_etal_2013,Petigura_etal_2013}, it is at the moment completely unclear whether the phenomenon of life is widespread in the universe or unique to our home planet. Therefore, considerable effort has been undertaken to identify suitable signatures of bioactivity (biosignatures) on exoplanets, in parallel to the search for extraterrestrial life within the Solar System (e.g.\ on Mars or Europa).

A promising approach to identify a biosphere is to use the influence of life on the composition of the atmosphere \citep[e.g.][]{Kaltenegger_etal_2010,Seager_Deming_2010,Rugheimer_etal_2013,Seager_2014}. For example, the detection of species out of chemical equilibrium has been argued to be indicative of a biosphere. This idea was first discussed by \citet{Lederberg_1965}, \citet{Lovelock_1965,Lovelock_1975} and \citet{Hitchcock_Lovelock_1967}. These authors argued that states out of thermodynamic equilibrium can be seen as a generalised signature of life. For instance, in Earth's atmosphere, oxygen and methane are severely out of redox equilibrium because of biological forcing \citep[e.g.][]{Sagan_etal_1993}. Without life, methane would indeed rapidly be removed from the atmosphere by reacting with oxygen.

The atmospheric properties of exoplanets as small as Earth have already been constrained \citep{deWit_etal_2016,Southworth_etal_2017}. However, measurements of the atmospheric composition or mass are not yet feasible. In the future, investigating the atmospheres of Earth analogues will become possible by using telescope concepts similar to \textit{Darwin} or the \textit{Terrestrial Planet Finder Interferometer} (TPF-I). These instruments are designed to directly image Earth-sized exoplanets in the habitable zone. Besides the spectroscopic characterisation of the atmosphere, they can also be used to study the surface reflectance \citep[e.g.][]{Hegde_Kaltenegger_2013}, including surface reflectance biosignatures \citep[e.g.][]{DesMarais_etal_2002}, for example from vegetation and also microbial mats \citep[e.g.][]{Seager_etal_2005,Hegde_etal_2015,Schwieterman_etal_2015}.

In general, the best strategy to avoid biosignature false positives \citep[e.g.][]{Rein_etal_2014} is to characterise the potentially inhabited exoplanetary system (including the host star) as accurately as possible. Only when seen in context can the significance of a biosignature detection be assessed. The study of methods yielding information in addition to atmospheric mass and composition, or general bulk properties (exoplanet radius, bulk density, orbital period), is clearly warranted. In particular, ways to probe the geology of exoplanets or detect non-atmospheric biosignatures would be valuable complements.

In this work, we investigate the idea of characterising an exoplanet by observing dust generated during an impact event. The composition of the dust holds information on the geology of the impacted planet and might contain biosignatures if the planet is inhabited. Depending on the impact parameters, the escaping debris can have a much larger surface area than the planet itself. Impact ejecta most easily escape from low gravity bodies. Thus, the largest amounts of dust are expected for small planets (or moons), the atmospheres of which are the most difficult to study.

Several studies have considered dust generated from collisions or impacts, covering different regimes in terms of sizes of both the impactor and the target. Giant collisions involving two bodies of planetary size are expected to occur frequently during the final stage of terrestrial planet formation lasting for approximately 100\,Myr \citep[e.g.][]{Kenyon_Bromley_2006,Kokubo_Genda_2010}. For example, \citet{Jackson_Wyatt_2012} modelled the evolution and detectability of dust originating from the Moon-forming collision. \citet{Genda_etal_2015} estimated the total amount of dust produced from giant impacts and compared to observations of warm debris disks (i.e.\ dust in the terrestrial region). \citet{Jackson_etal_2014} modelled observational signatures of giant impacts occurring at large orbital radii. \citet{Morlok_etal_2014} linked infrared observations of dust from collisions to laboratory spectra of terrestrial and martian rocks.

Dust can also be produced from mutual collisions of smaller bodies. This is commonly observed in the form of dusty debris disks, objects akin to the asteroid belt or the Kuiper belt in the solar system \citep[e.g.][]{Wyatt_2008}. In general, debris disk dust is believed to originate from a so-called collisional cascade: asteroidal or cometary bodies\footnote{These objects can be seen as leftover planetesimals that were not incorporated into planets.} collide and produce smaller bodies, which further collide to produce even smaller fragments, resulting in copious amounts of micron-sized dust grains. Analysing the dust properties allows for characterisation of the parent bodies at the top of the collisional cascade \citep[e.g.][]{deVries_etal_2012}. Collisions among asteroids or comets in debris disks can also be responsible for observed dust clumps \citep{Wyatt_Dent_2002,Kenyon_etal_2014}.

In this work, we are concerned with yet another impact regime: we consider impacts of asteroidal or cometary bodies, typically tens of kilometres in size, with planetary bodies. An example for such an event on Earth is the famous Chicxulub impact that caused the Cretaceous-Paleogene (K-Pg) extinction about 65\,Myr ago. In contrast to mutual collisions of asteroids or comets, such events generate dust originating from our objects of interest, namely exoplanets. In addition, such impacts are expected to occur over the entire lifetime of an exoplanetary system and are not restricted to the first 100\,Myr. In particular, they may occur once a planet has been extensively modified by geological activities or the presence of life, the signatures of which could be imprinted in the ejected dust.

We emphasise that the present work does not attempt to develop a detailed model covering all the subtleties of the impact process and the dust evolution. Rather, the intent of the paper is to present the idea and make some quantitative estimates. We employ a simplified approach to get a general idea of the ejected dust masses, the timescales on which the dust evolves and what instruments would be needed to observe impact-generated dust.

Our paper is organised as follows: section \ref{Sec:interesting_substances} is a general discussion of potentially interesting substances escaping during an impact event. Section \ref{Sec:Modelling} presents the modelling of the impact and the subsequent collisional evolution. In section \ref{Sec:Discussion} we discuss our results and estimate instrument capabilities needed to detect minerals or biological matter. Section \ref{Sec:Conclusions} gives a summary and the conclusions.

\section{Substances of interest escaping during an impact event}\label{Sec:interesting_substances}
\subsection{Minerals and rocks}
An obvious component of the escaping ejecta will be minerals and rocks from the planetary surface. While a large fraction of the escaping rocky material will be vaporised or molten and subsequently recondense, some percentage can actually remain in the solid state when transported from the planetary surface into orbit (section \ref{subsec:nondamaged}). Minerals and rocks can inform us about the geology of the exoplanet, but they can also carry information about a biosphere (if present), since organisms can interact with minerals in various ways \citep[e.g.][]{Banfield_etal_2001}. With more than 4000 known minerals, the mineralogical diversity of Earth is large compared to Venus or Mars \citep{Hazen_etal_2008}. The latter are expected to have at most a few hundred minerals, although they accreted from the same material as Earth during the planet formation process. In addition, Earth and Venus are almost identical in size. The difference in mineralogical diversity is thought to be due, directly or indirectly, to the presence of life on Earth \citep{Hazen_etal_2008}. Minerals that owe their existence to biology can therefore be considered a candidate biosignature. Obviously, a thorough understanding of the different (potentially abiotic) production pathways of a mineral is necessary to avoid false positives.

Life influenced Earth's mineralogy in different ways, e.g.\ by changing the chemistry of the oceans and the atmosphere, which allowed new mineral species to arise. \citet{Hazen_etal_2008} argued that most of the biologically mediated diversification of Earth's mineralogy took part from 2.5\,Gyr ago until today. For example, the rise in atmospheric oxygen due to photosynthesis between $\sim$2.2 to 2.0\,Gyr ago ("Great Oxygenation Event") resulted in a large number of new minerals produced by weathering of other minerals in an oxygen-rich environment \citep{Hazen_etal_2008}. Skeletal biomineralization became important in the Phanerozoic Eon 0.542\,Gyr ago and continues today. It is an example of direct production of minerals by living organisms. Major skeletal minerals are calcite, aragonite, magnesian calcite, apatite and opal. In terms of volume, the most important biominerals are calcium carbonates such as calcite. Their values as biosignature is limited by the fact that abiotic production pathways exist. However, carbonates are in general the result of aqueous processes. Their presence on a planet can therefore hold important clues about past or present habitability. For example, the detection of calcite on Mars has been interpreted as evidence for the presence of liquid water in the past \citep{Boynton_etal_2009}.

\citet{Rosing_etal_2006} suggested that granitic continents are signatures for (oxygenic or anoxygenic) photosynthetic life. Granitoid rocks\footnote{Granitoid rocks are predominantly composed of feldspar and quartz. Granite sensu stricto ("true granite") is one example of a granitoid rock. Other examples include tonalities, monzonites or quartz diorites.} (also called granites sensu lato; simply granite hereafter) are formed from the subduction and dehydration of altered basalts and sediments, and thus rely on weathering processes. Microbes can increase the rate of silicate weathering significantly. Using thermodynamic arguments, \citet{Rosing_etal_2006} concluded that only photosynthetic life is able to increase weathering rates sufficiently, thus stimulating the production of granitoids which led to the stabilisation of the continents in the early Archean. This might explain the temporal correlation between the possible emergence of photosynthesis and the stabilisation of the continents. It would also explain why granite is ubiquitous on the surface of life-bearing Earth while apparently absent on other planets, although tentative evidence exists for granitic terrains on Venus \citep{Hashimoto_etal_2008}. \citet{Bonin_2012} argued that granite should occur on any silicate planetary body, the question being only whether enough granite is build up to constitute stable continents.

An interesting variation of the above discussion concerns the role of minerals in the origin of life. Some scenarios of the origin of life rely explicitly on the presence of certain minerals, for example clays or transition metal sulfides, some of which are rather exotic \citep[e.g.][and references therein]{Hazen_etal_2008}. The detection of such minerals would then suggest that the mineralogy of the impacted planet has evolved sufficiently to support an origin of life \citep{Hazen_etal_2008}.

\subsection{Water}
Liquid water is essential to all life forms on Earth and is thought to be a prerequisite for life as we know it. Detecting exoplanets hosting liquid water (be it on the surface or e.g.\ beneath an icy crust) is therefore considered an essential step in the quest for extraterrestrial life. If an impactor hits an ocean or an ice crust, some water could escape the exoplanet and potentially be detectable in the form of water gas or water ice spectral features \citep[e.g.][]{Lisse_etal_2012}. More than 70\% of Earth's surface is covered by water so that oceanic impacts are actually more probable than land impacts. Some exoplanets might even be completely covered by oceans \citep[so-called waterworlds, e.g.][]{Berta_etal_2012}. However, it should be noted that the impactor itself might contain substantial amounts of water and could consequently pollute the ejecta. Also, the lifetime of water in the terrestrial region of a Sun-like star is very short, making a detection extremely improbable (section \ref{subsec:oceanic_impacts}).

\subsection{Direct detection of biological matter}
Direct evidence for a biosphere would come from the detection of e.g.\ microorganisms escaping during the impact. Microbes make up most of the biomass on Earth \citep{Madigan_2012}. Also, microbes have been present over a large fraction of Earth's history, in contrast to multicellular organisms. In addition, microbes are expected to dominate Earth's future biosphere when the Sun enters the late stages of its main-sequence evolution \citep{OMalley-James_etal_2013,OMalley-James_etal_2014}. Therefore, we argue that microbes are an attractive option if biological matter is to be detect directly.

Various authors have discussed the possibility of transferring microbes between planetary bodies with the help of rocks ejected during impact events \citep[e.g.][]{Melosh_1988,Mileikowsky_etal_2000,Worth_etal_2013}, an idea commonly known as lithopanspermia. We can use the knowledge acquired in these studies to investigate the possibility of a direct detection of microbial life ejected during an impact.

For remote detection, a number of authors have considered spectral signatures of microorganisms. \citet{Dalton_etal_2003} presented near-infrared laboratory spectra of three different microbial species and compared them to observations of Europa's surface from the \textit{Galileo} spacecraft. The most prominent spectral features are asymmetric absorption features due to water of hydration. The work by \citet{Dalton_etal_2003} was substantially extended by \citet{Hegde_etal_2015}. They measured the spectral characteristics of 137 different microorganisms over the visible and near-infrared wavelength range. \citet{Schwieterman_etal_2015} measured the spectral properties of microorganisms with non-photosynthetic pigments. \citet{Knacke_2003} discussed the prospects for detecting algae on the surface of exoplanets. We discuss the possibility to detect spectral features of microorganism ejected during an impact in section \ref{subsubsec:microorganism_detection}.

All known life on Earth exhibits an interesting property known as \emph{homochirality}: only left-handed L-amino acids are found in proteins, and only right-handed D-sugars are found in nucleic acids. As a result, optical activity can be induced by the interaction of light with the chiral biomolecules. This can lead a signal in the form of circular polarisation. A particularly strong interaction occurs for photosynthetic organisms. A series of papers investigated the potential of chiral signatures as biosignatures. General considerations were presented by \citet{Gledhill_etal_2007}. Laboratory measurements of circular polarisation signals from biological samples were published by \citet{Sparks_etal_2009} and \citet{Martin_etal_2010}. The idea was set into practice in a search of Mars' surface for circular polarisation signatures \citep{Sparks_etal_2005,Sparks_etal_2012}.

In a recent paper, \citet{Berdyugina_etal_2016} argued that linear, instead of circular, polarisation would be a more sensitive biosignature. The linear polarisation signatures are associated with absorption of light by various biopigments that are used by plants and bacteria for photosynthesis or protection. \citet{Berdyugina_etal_2016} presented laboratory measurements of reflectance and linear polarisation spectra for a number of plants, complemented with non-biological samples. These data were then used to calculate the linearly polarised spectra of Earth-like planets.

The results of all these different studies are indeed primarily directed towards modelling of disk-integrated spectra of exoplanets that may become available from future instruments capable to directly image Earth analogues. How does the situation change if we consider microorganisms ejected by an impact? First, interference from the atmosphere or clouds \citep[e.g.][]{Schwieterman_etal_2015,Berdyugina_etal_2016} is no longer an issue. Second, the ejected debris can represent a significantly larger surface area than the planet itself (see figure \ref{fig:frac_lum}). On the other hand, because of high temperature and pressure, a large fraction of any microorganisms would be destroyed during the impact event. Also, the density of microbes in debris originating from relatively deep below the planetary surface might be low, especially if photosynthetic organisms are considered. This essentially depends on how the depth of the biosphere compares to the depth of origin of the escaping ejecta (see section \ref{sec:excavation_depth}). In addition, the question about the stability of the molecules giving rise to spectral features under space conditions (vacuum, intensive radiation environment) arises. \citet{Dalton_etal_2003} gave some estimates based on the strength of molecular bonds. Experimental studies of the survival of microbes under space conditions \citep[e.g.][]{Mastascusa_etal_2014} can also give some useful hints. There exist also experiments investigating the integrity of biomolecules under space conditions directly \citep[e.g.][]{Lyon_etal_2010}. For chiral signatures, the racemisation timescale is an important additional quantity to consider.

\section{Modelling of the impact event and the collisional evolution of the ejecta}\label{Sec:Modelling}
When assessing the ejection of exoplanetary material into a circumstellar orbit by an impact event, different issues have to be considered. The amount of ejected material depends on the impactor's size, its speed, but also on the planet's mass. Certain substances may be damaged or completely destroyed by high pressure, melting or evaporation. It is also important to estimate the depth from which the ejected debris originate. In addition, once in orbit, the debris can collide and produce new, smaller debris. It is this collisional evolution that will determine the overall lifetime of the debris cloud and its luminosity.

We neglect the effect of the impacted body's atmosphere on both the impactor and the escaping material. Concerning the impactor, we may use the criterion developed by \citet{Shuvalov_etal_2014} to assess whether a given impactor is actually crater-forming, i.e.\ does not experience fragmentation and deceleration during the traversal of the atmosphere. \citet{Shuvalov_etal_2014} propose that a crater-forming impactor satisfies the following condition:
\begin{equation}\label{eq:impactor_atmosphere_criterion}
\frac{B_\mathrm{fr}a}{H}\sqrt{\frac{\rho_\mathrm{i}}{\rho_\mathrm{atm}}}>1
\end{equation}
with $B_\mathrm{fr}$ a constant equal to 3.5, $a$ the radius of the impactor, $H$ the scale height of the atmosphere, $\rho_\mathrm{i}$ the impactor's density and $\rho_\mathrm{atm}$ the atmospheric density at the surface of the exoplanet. Equation \ref{eq:impactor_atmosphere_criterion} implies that for a Venus-like atmosphere, a 1\,km-diameter asteroid or a 2\,km-diameter comet are crater-forming. For the atmosphere studied by \citet{Shuvalov_etal_2014} (a 200\,bar atmosphere with a scale height of 40\,km), a 8\,km-diameter asteroid or a 14\,km-diameter comet are crater-forming. In the following sections, we will consider impactors with a diameter of 20\,km that should thus be crater-forming even for thick atmospheres.

An atmosphere may also decelerate ejected material and thus hamper the escape of planetary surface material. However, a large impact can essentially remove the atmosphere from the impact site (N.\ Artemieva 2015, private communication) and allow fragments of any size to escape the planet if they are launched with a speed higher than the escape speed. This is in sharp contrast to small-scale impacts, where atmospheric drag can prevent small fragments from escaping \citep{Artemieva_Ivanov_2004}.

We use the study by \citet{Shuvalov_2009} to estimate under which conditions atmospheric effects can be neglected. \citet{Shuvalov_2009} defines an dimension-less impact strength given by\footnote{Note that there is a typo in equation 2 (the definition of $\xi$) in \citet{Shuvalov_2009}, that was corrected in equation 2 of \citet{Shuvalov_etal_2014}.}
\begin{equation}
\xi = \frac{8a^3\rho_\mathrm{i}}{H^3\rho_\mathrm{atm}} \frac{v_\mathrm{i}^2-v_\mathrm{esc}^2}{v_\mathrm{esc}^2} \frac{\rho_\mathrm{t}}{\rho_\mathrm{t}+\rho_\mathrm{i}}
\end{equation}
where $v_\mathrm{i}$ is the speed of the impactor, $v_\mathrm{esc}$ the escape speed and $\rho_\mathrm{t}$ the density of the target rock. From equation 7 in \citet{Shuvalov_2009}, we find that for $\xi>1000$, effects of the atmosphere on the escaping target material can be neglected\footnote{There are two typos in equation 7 of \citet{Shuvalov_2009}. It should read $\log\chi_\mathrm{t}=\min(\log(0.02),-4+0.767\log\xi)$ (Shuvalov 2016, private communication).}. Using the impactor speeds of the impact scenarios considered in section \ref{subsec:results}, we find that for an Earth-sized exoplanet with an Earth-like atmosphere, a 3--4\,km-diameter impactor is enough to neglect atmospheric effects. On the other hand, for a Venus-sized exoplanet with a Venus-like atmosphere, a $\sim$24\,km-diameter impactor is required. For smaller exoplanets, a thicker atmosphere is tolerable because the escape velocity decreases. We conclude that for the scenarios considered in this study (20\,km-diameter impactor), our approach to neglect the atmospheric effect on the escaping fragments is valid approximately up to atmospheres as thick as that of Venus.

\subsection{Amount of ejected material}\label{subsec:ejected_material}
We distinguish here between two estimates: the total amount of ejecta that can escape the target, and the amount of ejecta that can escape without being ``damaged'' by high shock pressure, that is, that can retain some detectable signature of ``pristine'' exoplanetary material (possibly biosignature).

\subsubsection{Total ejected mass}\label{subsec:totalejected}
To estimate the \emph{total} amount of escaping ejecta, we use an ejecta model calibrated by laboratory measurements \citep{Housen_Holsapple_2011}. For a rock to escape the planet's gravitation field, the ejection velocity $v_\mathrm{e}$ needs to be larger than $v_\mathrm{esc}=\sqrt{2GM/R}$ with $M$ and $R$ the planetary mass and radius and $G$ the gravitational constant. The mass ejected with a velocity larger than $v_\mathrm{e}$ can be estimated by
\begin{equation}\label{eq:total_ejecta}
M_\mathrm{e}(>v_\mathrm{e})=\frac{3k}{4\pi}\frac{\rho_\mathrm{t}}{\rho_\mathrm{i}}\left[\left(\frac{x(v_\mathrm{e})}{a} \right)^3-n_1^3\right] m
\end{equation}
Here $m$ is the mass of the impactor and $k=0.3$ and $n_1=1.2$ are constants \citep[table 3, rock target, in][]{Housen_Holsapple_2011}. For the target rock density $\rho_\mathrm{t}$, we assume 3000\,kg\,m$^{-3}$. The distance to the impact centre $x(v_\mathrm{e})$ can be computed numerically for a given ejection velocity $v_\mathrm{e}$ by using
\begin{equation}
v_\mathrm{e}=C_1\left[\frac{x}{a} \left(\frac{\rho_\mathrm{t}}{\rho_\mathrm{i}}\right)^\nu \right]^{-1/\mu}\left(1-\frac{x}{n_2R} \right)^p v_\mathrm{i}
\end{equation}
with $C_1=1.5$, $\nu=0.4$, $\mu=0.55$, $n_2=1.5$ and $p=0.5$ constants. Finally, the transient crater radius $R$ (in the gravity regime) is given by
\begin{equation}\label{eq:crater_radius}
R=H_1\left( \frac{m}{\rho_\mathrm{t}}\right)^{1/3}\left(\frac{\rho_\mathrm{t}}{\rho_\mathrm{i}} \right)^{(2+\mu-6\nu)/[3(2+\mu)]}\left(\frac{ga}{v_\mathrm{i}^2} \right)^{-\mu/(2+\mu)}
\end{equation}
where $g$ is the surface gravity of the target body and $H_1$ is a constant. Table \ref{tab:impact_results} lists transient crater radii for the impact scenarios considered in this work. Unfortunately, \citet{Housen_Holsapple_2011} do not give an estimate for $H_1$ in the case of a rocky target. We adopt $H_1=0.6$ based on values for other target materials, but note that our results only weakly depend on the exact value of $H_1$.

\subsubsection{Ejected mass of ``intact'' material}\label{subsec:nondamaged}
We now estimate the amount of material escaping an exoplanet without being affected by high pressure induced shock damage, melting or evaporation. This material will be most representative of the exoplanet because it was least modified by the impact. A lot of parallels can be drawn to the study of lithopanspermia, i.e.\ the study of possible interplanetary life transfer by means of ejected rocks. To avoid sterilisation of the rock during ejection, the temperature of the rock must not rise above $T_\mathrm{max}\sim370$\,K. In our case, the maximum tolerable temperature depends on the nature of the substance of interest. For minerals, $T_\mathrm{max}$ could be significantly higher and for example given by the temperature for melting. Interestingly, impact models predict that, while most ejecta reaching escape velocity melt or evaporate, a small fraction is in fact subject to low pressure and consequently low temperature. This is possible because the interference of the stress and rarefaction waves creates a region of low pressure, but with high pressure gradient that can accelerate fragments to escape velocity \citep{Melosh_1985}. The mass of these spalled ejecta can be estimated by using a formula originally derived by \citet{Melosh_1985}, with a mathematical error corrected by \citet{Armstrong_etal_2002}:
\begin{equation}\label{eq:spallation}
M_\mathrm{e}(>v_\mathrm{e},<P_\mathrm{max})=\frac{3P_\mathrm{max}}{4\rho_\mathrm{t}c_\mathrm{L}v_\mathrm{i}}\left[\left(\frac{v_\mathrm{i}}{2v_\mathrm{e}} \right)^{5/3}-1 \right]m
\end{equation}
where $M_\mathrm{e}(>v_\mathrm{e},<P_\mathrm{max})$ is the mass of ejecta leaving the surface with a velocity larger than $v_\mathrm{e}$ and shocked to a pressure equal or below $P_\mathrm{max}$ and $c_\mathrm{L}$ is the sound speed in the target rock ($\sim$6\,km\,s$^{-1}$). We see from equation \ref{eq:spallation} that the impactor's speed $v_\mathrm{i}$ needs to be larger than 2$v_\mathrm{esc}$ for any spalled rocks to leave the planet. On the other hand, there is also a maximum velocity a spalled fragment can be ejected with: $v_\mathrm{e,max}=v_\mathrm{i}/2$. Obviously equation \ref{eq:spallation} is only valid up to a certain value of $P_\mathrm{max}$ \citep[see][footnote 4, for the constraint on $P_\mathrm{max}$ to be fulfilled for applicability of the equation]{Melosh_1985}. We note that the spallation theory leading to equation \ref{eq:spallation} is in principle only valid for ejection velocities up to 1\,km/s \citep{Melosh_1984}, but has been used for larger ejection velocities by various authors \citep[e.g.][]{Mileikowsky_etal_2000,Armstrong_etal_2002}, including \citet{Melosh_1985} himself. We show in section \ref{subsec:compatibility} that it provides reasonable estimates of the spalled mass when compared to numerical simulations.

Depending on the nature of the substance under consideration, $P_\mathrm{max}$ takes different values. In the case of microorganisms, \citet{Mileikowsky_etal_2000} adopted a maximum acceptable temperature of 370\,K, corresponding to $P_\mathrm{max}=1$\,GPa as the limiting pressure for survival. However, \citet{Artemieva_Ivanov_2004} argued that launch of Martian meteorites is not possible without substantial ($>10$\,GPa) compression, but that the temperature increase may still be well below 100\,K. For our study, the survival of the ejected microbes is not important. Rather, they should remain intact to a degree so to retain their spectral characteristics making them detectable. Therefore, we adopt $P_\mathrm{max}=10$\,GPa for microbes. For minerals on the other hand, we set $P_\mathrm{max}$ equal to 50\,GPa. This can be considered a typical, though conservative\footnote{Conservative in the sense that we are not likely overestimating the fraction of undamaged material.} pressure where rocks or minerals start to melt. For example, according to \citet{Pierazzo_Melosh_1999}, the pressure for incipient melting is 46\,GPa for granite and 135\,GPa for dunite. \citet{Melosh_1989} lists pressures of incipient melting for limestone (66\,GPa, but calcite decarbonation starts already at 45\,GPa) and granite (78\,GPa, i.e.\ different from the \citet{Pierazzo_Melosh_1999} value). For granite and dunite, the pressure for complete melting is only slightly above the one for incipient melting \citep{Pierazzo_Melosh_1999}. $P_\mathrm{max}=50$\,GPa is used to determine the fraction of the ejecta escaping without melting or evaporation. Fragments produced from melt or vapour have a different size distribution which affects their collisional evolution (sections \ref{size_dist} and \ref{coll_evolution}).

\subsubsection{Compatibility of the two prescriptions}\label{subsec:compatibility}
Our choice of using two different equations for the spalled ejecta that are subject to a maximum pressure constraint and the total ejecta is in principle not self-consistent. However, we can check that they do not give incompatible results for the range of parameters considered here. One good test is to check if equation \ref{eq:spallation} tends towards equation \ref{eq:total_ejecta} when $P_\mathrm{max}$ is set to the maximum value it can reach anywhere within the target. We follow \citet{Melosh_1985} who found that this maximum pressure should be of the order of a few hundred GPa for a violent impact event, and consider $P_\mathrm{max}=400$\,GPa. Putting this value in equation \ref{eq:spallation} (although the equation might not be applicable anymore for such a high value of $P_\mathrm{max}$), we find a mass of spalled ejecta that agrees within a factor $\sim$3 with the total ejecta given by equation \ref{eq:total_ejecta}.
Another check is to compare predictions of equations \ref{eq:spallation} and \ref{eq:total_ejecta} to estimates obtained from detailed numerical simulations of impact events. Taking as a first reference the results of \citet{Artemieva_Morgan_2009}, who simulated the Chicxulub impact and calculated both the amount of solid escaping ejecta and total escaping ejecta, we find an order of magnitude agreement with both our spalled ejecta estimate (equation \ref{eq:spallation}) with $P_\mathrm{max}=46$\,GPa (no melting) and the total ejecta estimate from equation \ref{eq:total_ejecta}. 
As a second test, we take the results of \citet{Artemieva_Ivanov_2004}, who numerically studied the ejection of Martian meteorites, defined as ejecta with $P_\mathrm{max}<50$\,GPa. The corresponding prediction of equation \ref{eq:spallation} agrees with the simulation results within a factor of $\sim$2. We then consider the results of \citet{Artemieva_Shuvalov_2008}, who simulated impact ejecta escaping the Moon for both asteroidal and cometary impactors and kept track of the maximum pressure experienced by the ejecta. Comparing with predictions from equations \ref{eq:spallation} and \ref{eq:total_ejecta}, we again find agreement within a factor of $\sim$3 for the spalled ejecta and a factor better than 2 for the total ejecta. Finally, we compare our calculations to simulations of impacts on Titan by \citet{Artemieva_Lunine_2005} and \citet{Korycansky_Zahnle_2011} with predictions from equation \ref{eq:total_ejecta} and again find agreement within a factor of $\sim$3. In summary, we conclude that the usage of equations \ref{eq:spallation} and \ref{eq:total_ejecta} is justified for our order-of-magnitude study. 

\subsubsection{Results}\label{subsec:results}
Table \ref{tab:impact_results} shows the total escaping mass (both in absolute units and relative to the impactor mass) as well as the fraction subject to peak pressures below $P_\mathrm{max}$ for the different impact scenarios considered in this study. The impactor's diameter is fixed to 20\,km (for comparison, the diameter of the K-T impactor is estimated to $\sim$10\,km). Note that the ejecta \emph{fraction} subject to a certain pressure constraint is independent of the impactor's mass (equation \ref{eq:spallation}).

We consider impacts on Earth, Mars and the Moon as proxies to represent exoplanets of different sizes. The size distribution of exoplanets smaller than Earth is poorly constrained at the moment, but current data are consistent with a distribution that continues to rise towards smaller radii \citep[e.g.][and references therein]{Morton_Swift_2014,Bovaird_etal_2015}. Exoplanets down to the size of Mercury and smaller have already been detected \citep{Barclay_etal_2013,Campante_etal_2015}.

We consider two kinds of impactors: asteroids and comets (i.e.\ impactors of different densities and impact velocities). We take the distribution of asteroid and comet impact velocities upon Earth \citep{Steel_1998,Jeffers_etal_2001} as a basis for the considered impact events, acknowledging that an exoplanetary system might have a different architecture resulting in different dynamics and a different impact velocity distribution. For asteroids, we assume an impact velocity upon Earth of $v_\mathrm{i,E}=33$\,km\,s$^{-1}$. Only a relatively small fraction ($\lesssim$10\%) of impacting asteroids have $v_\mathrm{i,E}>33$\,km\,s$^{-1}$ \citep{Jeffers_etal_2001}. However, no escaping spalled material would be generated for the Earth-sized planet if the mean asteroid impact velocity was taken (mean impact velocity for asteroids is 21.7\,km\,s$^{-1}$ while the escape velocity of the Earth is 11.2\,km\,s$^{-1}$, i.e.\ $v_\mathrm{i}\ngtr2v_\mathrm{esc}$). Comets, on the other hand, have much higher impact velocities. We assume an impact velocity upon Earth of $v_\mathrm{i,E}=65$\,km\,s$^{-1}$. From this, we assign impact velocities upon the Mars- and Moon-like bodies by taking into account their lower masses and invoking energy conservation, i.e.\ $v_\mathrm{i,M}^2=v_\mathrm{i,E}^2-v_\mathrm{esc,E}^2+v_\mathrm{esc,M}^2$ (where the subscript M stands for Mars or Moon). For the density of the impactor, we assume  3000\,kg\,m$^{-3}$ for asteroids and $1000$\,kg\,m$^{-3}$ (water ice) for comets. Note that comets would lead to a dust signature from the target that is less "polluted" by impactor material, due to the comets higher volatile content compared to asteroids \citep{Morlok_etal_2014}.

\begin{table*}[p]
\caption{Impact velocities, total escaping masses and percentages subject to pressures lower than $P_\mathrm{max}=50$\,GPa respectively 10\,GPa for targets analogue to Earth, Mars and the Moon. The impactor's diameter is 20\,km (corresponding to $m=1.3\times10^{16}$\,kg for the asteroid and $m=4.2\times10^{15}$\,kg for the comet). Cometary impactors have higher impact velocity, but lower density compared to asteroids. The table also shows values for the transient crater diameter $R$, the excavation depth $H_\mathrm{exc}$ and the spall thickness $z_\mathrm{spa}$.}
\label{tab:impact_results} 
\centering 
\begin{tabular}{c c | c c c c c c} 
\hline\hline 
parameter & & \multicolumn{3}{c}{asteroid} & \multicolumn{3}{c}{comet} \\
& & Earth & Mars & Moon & Earth & Mars & Moon\\
\hline
$v_\mathrm{i}$ & km\,s$^{-1}$ & 33.0 & 31.5 & 31.1 & 65.0 & 64.2 & 64.1 \\
$M(>v_\mathrm{esc})$ & $10^{15}$\,kg & 7.2& 27.2 & 89.8 & 5.6 & 23.5 & 79.1 \\
$M(>v_\mathrm{esc})/m$& & 0.6 & 2.2 & 7.1 & 1.3 & 5.6 & 18.9\\
$<10$\,GPa & \% & 2.0 & 3.5 & 4.1 & 2.3 & 2.4 & 2.6\\
$<50$\,GPa (non-molten) & \%& 10.1 & 17.4 & 20.5 & 11.7 & 12.1 & 13.0\\
molten or vapourised & \%& 89.9 & 82.6 & 79.5 & 88.3 & 87.9 & 87.0\\
$R$ & km & 72.1 & 87.0 & 103.6 & 68.4 & 83.9 & 100.2 \\
$H_\mathrm{exc}$ & km & 14.4 & 17.4 & 20.7 & 13.7 & 16.8 & 20.0 \\
$z_\mathrm{spa}$ & m& $8.4$ & $15.5$ & $30.4$ & $7.0$ & $14.4$ & $29.6$ \\ 
\hline  
\end{tabular}
\end{table*}

\subsection{From what depth are escaping fragments originating?}\label{sec:excavation_depth}
In this section we estimate the depth below the exoplanetary surface from which escaping material can originate. This estimate needs to be related to e.g.\ the depth of the subsurface biosphere or the geological depth profile.

A first important parameter is the excavation depth, which is the maximum depth from which material is launched upwards above the target's surface after the impact, and thus represents an upper limit on the depth of origin of any ejecta. It does \emph{not} correspond to the full depth of the crater, because material below the maximum depth of excavation is displaced downwards instead \citep[see e.g.\ figures 5.13 and 5.14 in][]{Melosh_1989}. \citet{Melosh_1989} provided an estimate of the excavation depth $H_\mathrm{exc}$ from the transient crater radius $R$:
\begin{equation}
H_\mathrm{exc}\approx\frac{R}{5}
\end{equation}
The radius of the transient crater can be calculated from equation \ref{eq:crater_radius}. Thus, we can conservatively say that the depth of origin of the escaping ejecta is $z< H_\mathrm{exc}$. Table \ref{tab:impact_results} lists $H_\mathrm{exc}$ for the impact scenarios considered in this work. We find the excavation depth to be comparable to the impactor diameter.

However, in the main excavation flow the material ejected at high velocity near the impact point generally comes from shallower levels in the target. Given typical Maxwell z-model particle trajectories \citep[e.g.][]{Maxwell_1977,Wada_etal_2004}, one can estimate that the ejecta launched with escape velocity originate from a depth that is at least a factor of a few smaller than the excavation depth \citep[see e.g.\ figure 3 of][for $x/R\sim0.3$, estimated from the model in section \ref{subsec:totalejected}]{Wada_etal_2004}. Given the values for $H_\mathrm{exc}$ listed in table \ref{tab:impact_results}, this means that escaping ejecta do not originate deeper than a few kilometres below the surface, i.e. comparable to the limits of the deep biosphere on Earth \citep[e.g.][]{Cockell_Barlow_2002}. Additionally, the \emph{low-pressure} (spalled) ejecta launched with high velocity are thought to originate from a zone even closer to the surface, characterised by the spall thickness $H_\mathrm{spa}$ \citep[denoted $z_\mathrm{S}$ by][]{Melosh_1985}. It can be estimated from the equation given by \citet{Melosh_1985}. As can be seen in Table \ref{tab:impact_results}, the spall thickness is of the order of only 10--30 metres.

\subsection{Size distribution of the escaping fragments}\label{size_dist}
The size distribution of the fragments is a fundamental property, as it will determine their subsequent collisional evolution (see section \ref{coll_evolution}).
We need here to clearly distinguish between the two populations of post-impact solid fragments: 
1) The ``spalled population'', i.e., fragments that escape in the solid state (spalled ejecta, typically a few percent up to 20\%, see table \ref{tab:impact_results}), and 2) the ``recondensed population'', corresponding to escaping fragments produced from the recondensation of melt or vapour.

\subsubsection{Size distribution profile}\label{size_distprof}
The size distribution of the produced fragments in the aftermath of violent impacts depends on several geometrical, dynamical and physical parameters and is difficult to constrain with precision, as revealed by several laboratory and numerical studies over the past three or four decades \citep[e.g.][]{Capaccioni_etal_1986,Housen_Holsapple_2011,Durda_etal_2015}. In most cases, however, this distribution can be approximated by a single power law of the form \citep[e.g.][]{Buhl_etal_2014}
\begin{equation}\label{PSD_powerlaw}
N(>D)\propto D^{-A}
\end{equation}
where $N(>D)$ is the number of particles with a diameter larger than $D$ and $A$ is the power law exponent. \citet{Buhl_etal_2014} compiled measurements of the exponent $A$ from impact experiments for different target materials. Most measurements yield $A\approx 2.5$ independent of target material. Interestingly, the same $A=2.5$ power law exponent is also theoretically expected for the size distribution arising from a collisional cascade in a steady state \citep{Dohnanyi_1969}. This significantly simplifies the modelling of the collisional evolution for both the spalled and recondensed material (see section \ref{coll_evolution}). For the spalled population, we can assume that the form of the size distribution remains constant, regardless of whether debris have been collisionally reprocessed or not\footnote{Poynting-Robertson drag, on the other hand, can change the shape of the size distribution.}. It also allows us to assume that particles recondensed from melt or vapour follow the size distribution of equation \ref{PSD_powerlaw} after a build-up phase where the initially monosized population is transformed into a steady-state collisional cascade.

Note that equation \ref{PSD_powerlaw} applies in principle to \emph{all} produced fragments and does not necessarily apply to \emph{escaping} fragments. It might overestimate the number of large escaping fragments, because large fragments tend to have lower velocities and have smaller chances to reach escape velocity \citep[as predicted by the Grady-Kipp distribution, see][]{Artemieva_Ivanov_2004}. However, given that the size-dependent velocity distribution is in itself poorly constrained and that the size distribution might also change due to unknown local conditions such as surface properties, water content of the target rock etc.\ we chose to keep equation \ref{PSD_powerlaw} as a reasonable first-order approximation of the \emph{initial} size distribution of the spalled population. The situation is radically different for the recondensed population. It is indeed expected that melt or vapour recondense into particles of a relatively narrow size range \citep{Johnson_Melosh_2012,Johnson_Melosh_2014}. As a consequence, we will here assume that all recondensed material has \emph{initially} a monosized distribution.

\subsubsection{Size range}\label{size_distrange}
To complete the description of escaping fragment sizes, the upper and lower cut-off of the size distribution need to be estimated. Regarding the lower end of the fragments size distribution, we can assume that it stretches all the way down to the stellar blowout size $D_\mathrm{bl}$, below which grains are expelled from the system by the radiation force of the host star. If we assume blackbody grains, this force is given by
\begin{equation}\label{eq:radiation_force}
F_\mathrm{rad}(D)=\frac{L_*D^2}{16cr^2}
\end{equation}
where $L_*$ is the stellar luminosity and $r$ is the distance between the grain and the star. Then, the ratio between radiation force and stellar gravity is then given by
\begin{equation}\label{eq:beta}
\beta(D)=\frac{F_\mathrm{rad}}{F_\mathrm{grav}}=\frac{3L_*}{8\pi cGM_*D\rho}
\end{equation}
with $M_*$ the stellar mass and $\rho$ the grain's density. If we consider that small grains are produced from parent bodies on circular Keplerian orbits, then the blowout size $D_\mathrm{bl}$ is obtained by setting $\beta=1/2$. For a solar type star, this is typically of the order of a micrometer.

Concerning the size of the largest fragment, we note that the two fragment populations do here strongly differ from one another, the spalled population extending up to fragments much larger than those of the recondensed population. The aforementioned studies by \citet{Johnson_Melosh_2012,Johnson_Melosh_2014} that investigated the recondensation of vapour into spherules as well as molten material into melt droplets and accretionary impact lapilii, have found that such particles have a narrow range of sizes typically peaking between 100\,$\mu$m to 1\,mm. We shall assume that all recondensed particles initially have a size equal to the expected value of the melt droplet size, which we calculated from equation (7) in \citet{Johnson_Melosh_2014} and the probability density distribution of the ejection velocity described in section \ref{coll_evolution_anal}. This initial particle size is obviously also the largest particle size within the recondensed population during its collisional evolution (collisions act exclusively destructively) and is listed as $D_\mathrm{max}$ in table \ref{tab:recondensed}. In fact, there will be a dispersion of particle sizes, but an initially monosized population is the simplest assumption and is closer to reality than a power law distribution extending down to the blowout size. Taking the melt droplet size as representative even for spherules produced from \emph{vapour} is justified since \citet{Johnson_Melosh_2014} found that the vapour plume is not dominating the high velocity ejecta. Also, melt droplet sizes are of the same order of magnitude as the spherule sizes computed by \citet{Johnson_Melosh_2012}.

As for the spalled population that was neither vaporised nor molten, we estimate the size of the largest fragment from \citep{Grady_Kipp_1980,Melosh_1985,Melosh_1992}\footnote{Note that there seems to be a typo in \citet{Artemieva_Ivanov_2004} for the expression of $D_\mathrm{max}$; their first term of the equation reads $(m_\mathrm{w}+3)/2$.}:
\begin{equation}\label{eq:Dmax}
D_\mathrm{max}=\frac{m_\mathrm{w}+2}{3}\frac{T}{\rho_\mathrm{t}v_\mathrm{e}^{2/3}v_\mathrm{i}^{4/3}}2a
\end{equation}
where $m_\mathrm{W}$ is a Weibull constant with the value 9.5 for basalt and $T$ is the tension at fracture (typically 0.1\,GPa for basalt and other igneous rocks). This equation is in principle derived for a  fragment size distribution (the Grady-Kipp distribution) that is different from the $A=2.5$ power law distribution, but we still use it as an order of magnitude estimate of the upper cut-off. We find that $D_\mathrm{max}$ is typically a few metres, depending on the impact and ejection velocities assumed (see table \ref{tab:spalled}).

\subsection{Collisional evolution of the fragment cloud}\label{coll_evolution}
Once in a circumstellar orbit, the ejected debris start to collide and produce new, smaller fragments. The detailed quantitative investigation of the fragment cloud's collisional evolution would be a long-term undertaking, requiring the use of advanced numerical tools \citep[like the new-generation LIDT-DD code by][]{Kral_etal_2013,Kral_etal_2015} which greatly exceeds the scope of the present study. We follow here instead a simplified analytical approach in the spirit of \citet{Wyatt_etal_2007} or \citet{Jackson_etal_2014} and determine the typical collisional timescales within the fragment population. We also assume for simplicity that the fragments are distributed within an axisymmetric circumstellar belt. This assumption is clearly erroneous in the first stages after the impact since the debris originate from a specific point in space. However, because of Keplerian shear, they will relatively quickly spread in longitude, forming a ring-like structure. This ring will be asymmetric at first, with all ejecta orbits passing through the initial impact location, but it will progressively become axisymmetric due to the target planet's perturbations \citep[see e.g.][]{Jackson_Wyatt_2012} and also mutual collisions amongst the debris \citep{Kral_etal_2015}.

We make an additional assumption, similar to that of \citet{Jackson_Wyatt_2012} for the post-Lunar-forming impact, and consider that the recondensed and spalled populations do not interact with each other and have separate collisional evolutions. We check the validity of this assumption with test runs using the statistical collisional code of \citet{Thebault_Augereau_2007}, showing that (for the reference asteroid-on-Earth impact case) the evolution of the spalled plus recondensed run is very close to the sum of the spalled-only and recondensed-only runs. The difference in terms of the total fractional luminosity is always less than 10\%.

\subsubsection{Analytical prescription}\label{coll_evolution_anal}
Following  \citet{Wyatt_Dent_2002} and \citet{Wyatt_etal_2007}, the collisional lifetime of a particle of diameter $D$ within a belt at distance $r$ from the host star and width $\Delta r$ can be expressed as \footnote{As underlined in the previous paragraph, the formalism of equation \ref{eq:t_coll} ignores the initial asymmetries in the ejecta's spatial distribution, but it is a reasonable order-of-magnitude approximation of collision lifetimes once the Keplerian shear and planet perturbations have morphed the fragment cloud into a ring-like structure. Note that the advanced simulations of \citet{Kral_etal_2015} have also confirmed that equation \ref{eq:t_coll} \citep[or its equivalent version found in][]{Loehne_etal_2008} is a satisfying first-order estimate of collision rates amongst post-impact debris}
\begin{equation}\label{eq:t_coll}
t_\mathrm{coll}(D) = \frac{2It_\mathrm{per}r\Delta r}{\sigma_\mathrm{c}(D)f(e,I)}
\end{equation}
where $I$ is the mean orbital inclination, $t_\mathrm{per}$ is the orbital period at $r$, $\sigma_\mathrm{c}(D)$ is the catastrophic cross-section (defined below) seen by a particle of diameter $D$ and $f(e,I)$ is the ratio between the relative velocity between the fragments $v_\mathrm{rel}$ and the Keplerian velocity $v_\mathrm{Kep}$ at $r$. The value of $f(e,I)$ depends on the inclination and the orbital eccentricity $e$. We can estimate the mean eccentricity by setting $\mean{e}=\mean{v_\infty}/v_\mathrm{Kep}$, where $\mean{v_\infty}$ is the mean speed of the escaping fragments once they have left the gravitational field of the planet (i.e.\ at ``infinite distance'' from the planet). We estimate $\mean{v_\infty}$ by first using equations \ref{eq:total_ejecta} and \ref{eq:spallation} to derive probability density distributions $p(v_\mathrm{e})$ of the ejection velocities\footnote{For simplicity, we use equation \ref{eq:total_ejecta} to compute the probability distribution of the recondensed population, although this equation describes the total ejecta. This is an acceptable simplification since recondensed material always makes up more than 80\% of the total escaping mass (see Table 1).}, i.e.\ $p(v_\mathrm{e})\mathrm{d}v_\mathrm{e}$ is the probability of an \emph{escaping} fragment to be ejected with $v_\mathrm{e}\pm\mathrm{d}v_\mathrm{e}$. This is achieved by taking the derivative of equations \ref{eq:total_ejecta} and \ref{eq:spallation} with respect to $v_\mathrm{e}$ and then normalising such that the integral from $v_\mathrm{esc}$ to the maximum ejection velocity is unity. By using equations \ref{eq:total_ejecta} and \ref{eq:spallation} to derive $p(v_\mathrm{e})$, we have assumed that the size distribution of a group of fragments with ejection velocity $v_\mathrm{e}$ does not depend on the value of $v_\mathrm{e}$. We transform $p(v_\mathrm{e})$ to a distribution $p(v_\infty)$ by using the relation $v_\mathrm{e}^2=v_\mathrm{esc}^2+v_\infty^2$. From $p(v_\infty)$, we calculate the expected value $\mean{v_\infty}$. The width of the ring can also be estimated from the mean eccentricity: $\Delta r\approx\mean{e}\cdot r$. In addition, we set $I=e/2$ by assuming equipartition between in-plane and out-of-plane velocities. 

\begin{figure}
\includegraphics[width=1\linewidth]{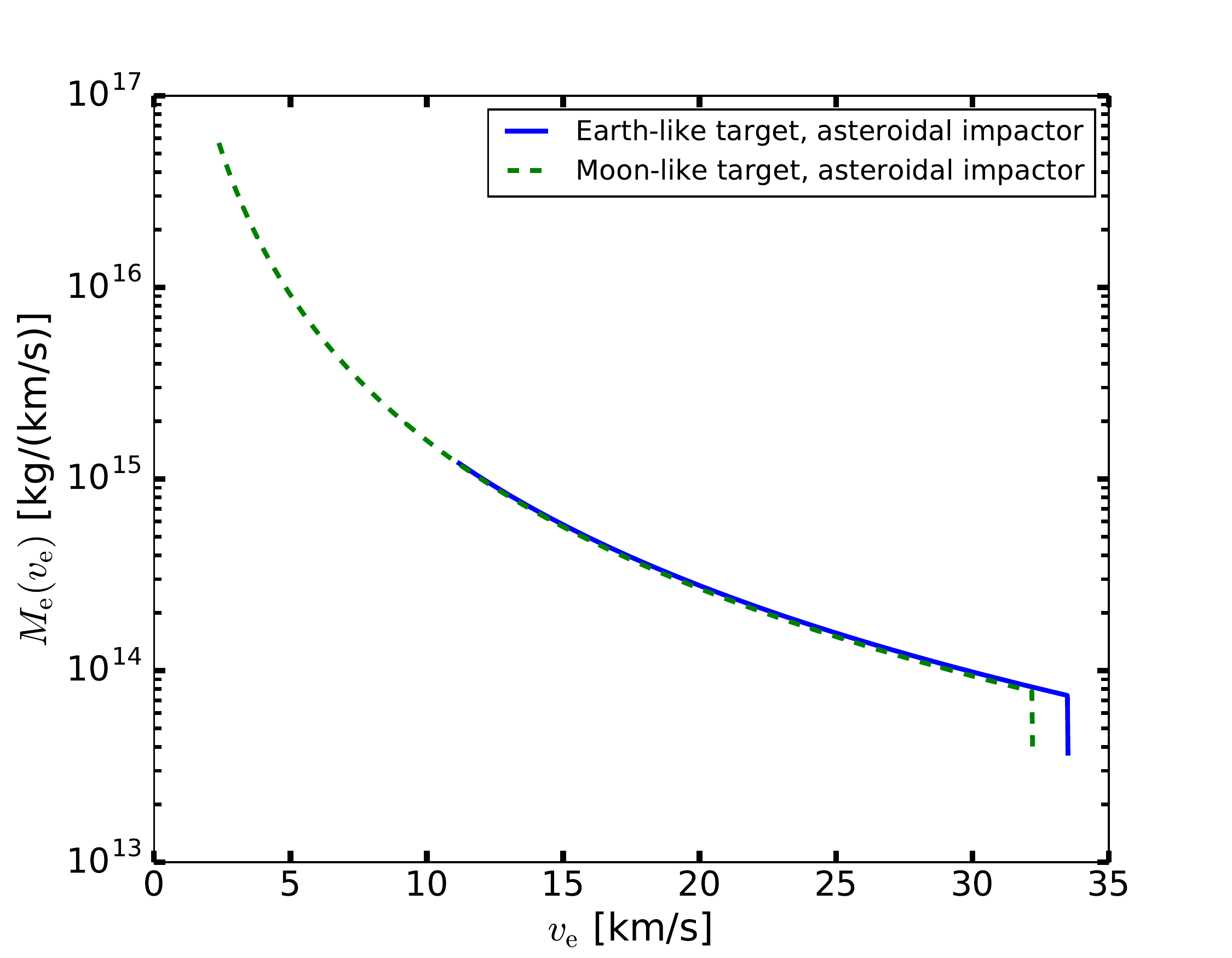}
\caption{Total escaping mass (in differential form, i.e.\ mass per ejection velocity bin $\mathrm{d}v_\mathrm{e}$) as a function of ejection velocity $v_\mathrm{e}$, for an asteroid impact, derived from equation \ref{eq:total_ejecta}. It can be seen that for a Moon-like target, the ejected mass is more dominated by fragments launched with a velocity just above the escape velocity (note that the y-axis is logarithmic). For the Earth-like target, the overall distribution is flatter, resulting in higher mean orbital velocity and consequently higher mean eccentricity.}
\label{fig:Me_ve}
\end{figure}

\begin{figure}
\includegraphics[width=1\linewidth]{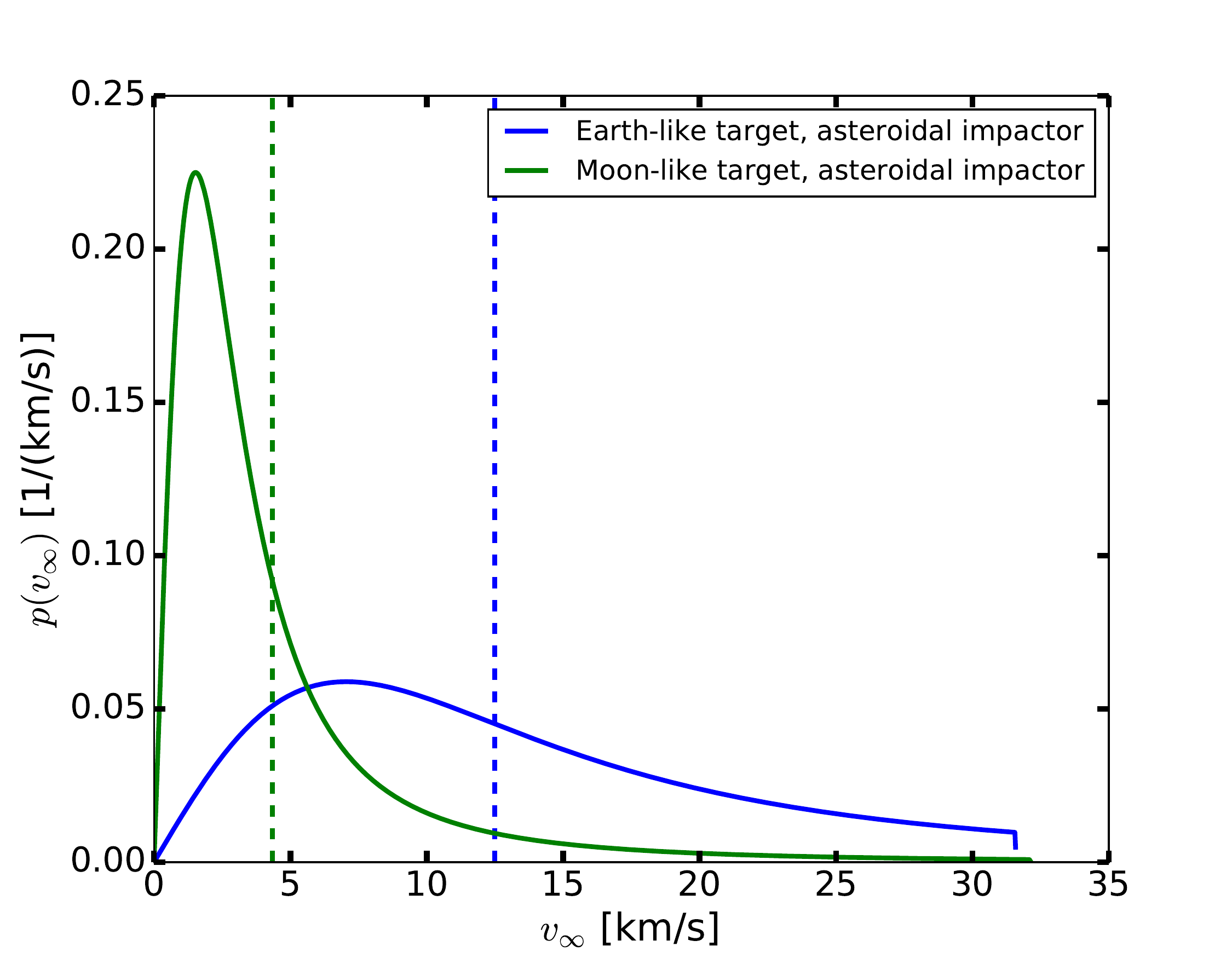}
\caption{Probability density distribution of $v_\infty=\sqrt{v_\mathrm{e}^2-v_\mathrm{esc}^2}$ for an asteroid impact onto an Earth- and Moon-like target, derived from equation \ref{eq:total_ejecta}. The vertical dashed lines indicate the expected values of the distributions.}
\label{fig:p_vinfty}
\end{figure}

The computed mean eccentricities $\mean{e}$ are given in tables \ref{tab:spalled} and \ref{tab:recondensed}. These results might appear counter-intuitive as, for the same impactor, we obtain higher $\mean{e}$  (and therefore higher $\mean{v_\infty}$) for the larger targets, for which the escaping fragments have to overcome a higher gravitational potential. This is due to the shape of the $M(v_\mathrm{e})\mathrm{d}v_\mathrm{e}$ curves of equations \ref{eq:total_ejecta} and \ref{eq:spallation}, which decrease steeply with $v_\mathrm{e}$ for low $v_\mathrm{e}$ values while having a flatter shape for higher ejecta velocities (see Figure \ref{fig:Me_ve}). Basically, the higher escape velocity of the Earth-like target truncates the distribution at a relatively high launch velocity, allowing only the tail to escape. Since the tail is relatively flat, the resulting $p(v_\infty)$ is rather flat as well, resulting in a higher $\mean{v_\infty}$. On the other hand, the total mass escaping from a Moon-like target is more dominated by fragments with $v_\mathrm{e}$ just above the escape velocity, resulting in low $\mean{v_\infty}$ and consequently low $\mean{e}$ (Figure \ref{fig:Me_ve}). This is also clearly seen in Figure \ref{fig:p_vinfty}, which shows the corresponding probability density distributions for $v_\infty$.

The catastrophic cross-section $\sigma_\mathrm{c}(D)$ depends on the minimum size $D_\mathrm{c}(D)$ a particle needs to have in order to collisionally destroy a particle of diameter $D$. It is given by $D_\mathrm{c}(D)=(2Q_D^*/v_\mathrm{rel}^2)^{1/3}D$, where $Q_D^*$ is the specific energy needed to break up the target. We describe the dependence of $Q_D^*$ on the target size $D$ with a broken power law with parameters determined by \citet{Benz_Asphaug_1999}, although the mean relative velocities we encounter in the case of Earth- or Mars-like targets are higher than what was considered by these authors. The catastrophic cross-section $\sigma_\mathrm{c}(D)$ seen by a fragment of diameter $D$ is then given by
\begin{equation}
\sigma_{c}(D)=\int_{D_\mathrm{x}(D)}^{D_\mathrm{max}}\frac{(D+D^\prime)^2}{4}\pi N(D^\prime)\mathrm{d}D^\prime
\end{equation}
with $N(D^\prime)$ the size distribution in differential form\footnote{This means that $N(D)\mathrm{d}D$ is the number of particles within an infinitesimal diameter bin $\mathrm{d}D$.} and $D_\mathrm{x}(D)=D_\mathrm{c}(D)$ if $D_\mathrm{c}(D)>D_\mathrm{min}$ and $D_\mathrm{x}(D)=D_\mathrm{min}$ otherwise ($D_\mathrm{min}$ is the minimum size present in the collisional cascade).

\subsubsection{Collisional vs. Poynting-Robertson timescales}\label{coll_evolution_cpr}
\begin{figure}
\includegraphics[width=1\linewidth]{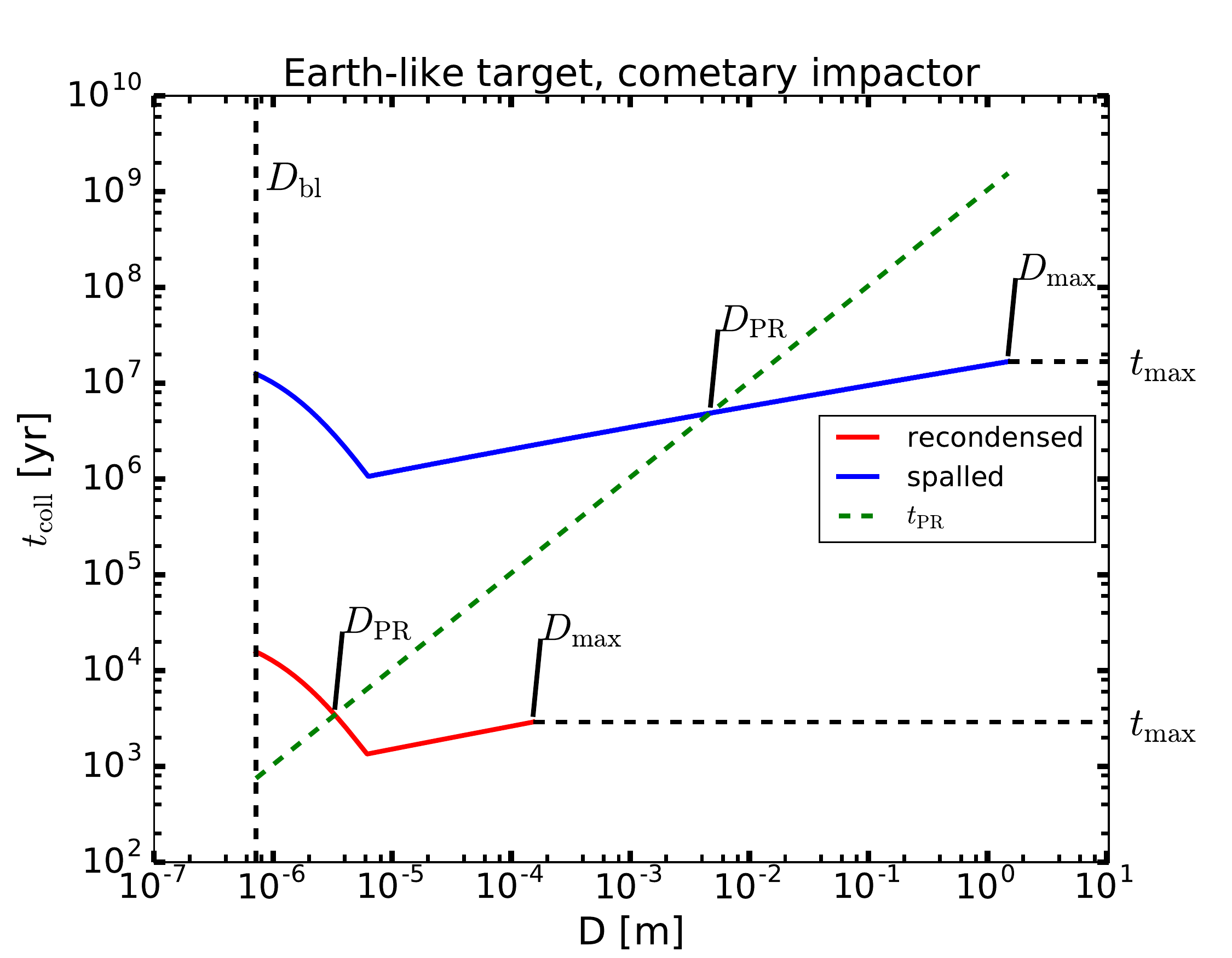}
\includegraphics[width=1\linewidth]{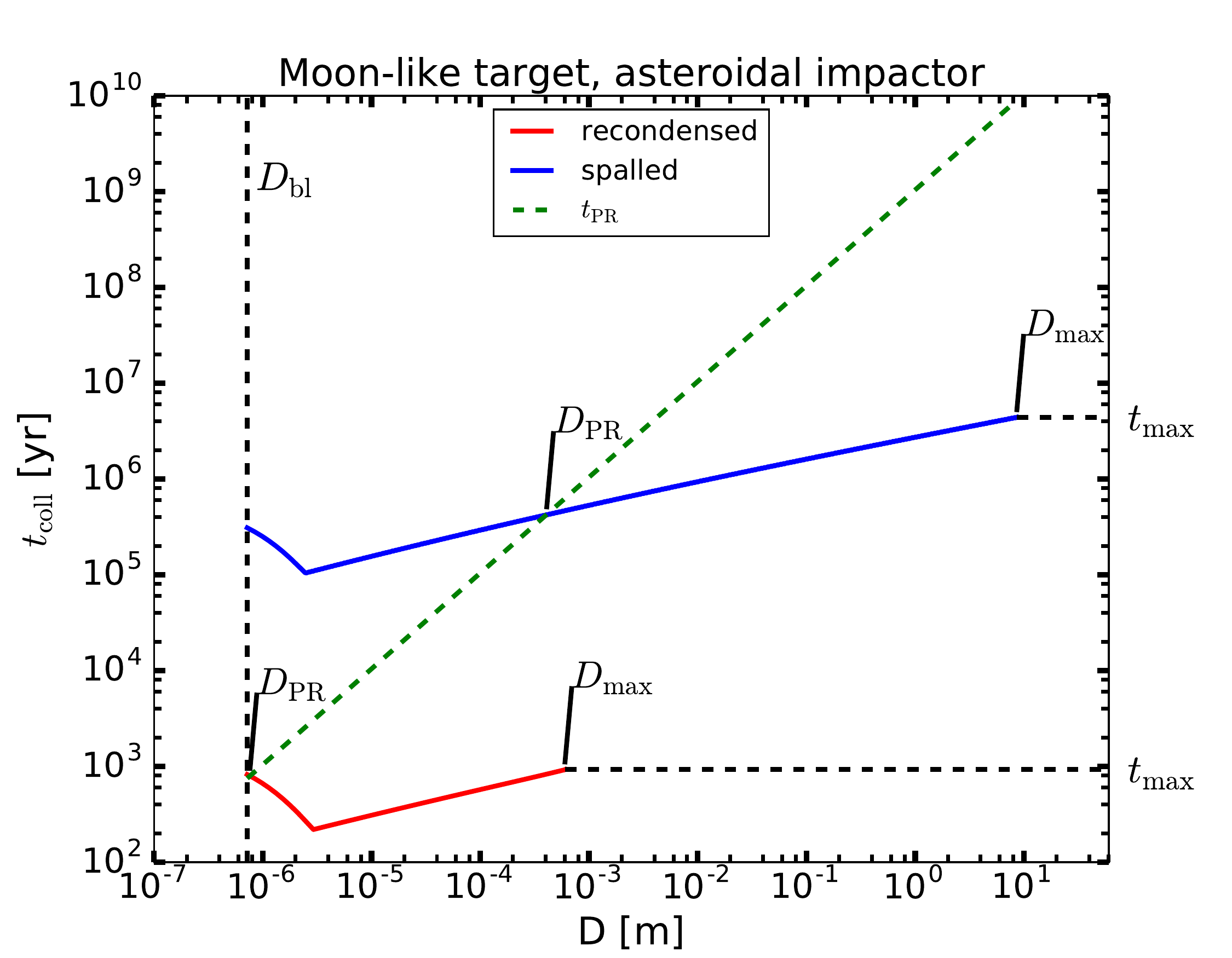}
\caption{Collisional timescales in the case of an Earth-like target impacted by a comet (top) and a Moon-like target impacted by an asteroid (bottom). The intersection of the collisional timescales and the timescale for PR-drag provides an estimate for the size $D_\mathrm{PR}$ below which there is a dearth of grains due to PR-drag. The collisional timescale of the largest fragments $t_\mathrm{max}$ sets the typical survival time of the post-ejection cloud.}
\label{fig:timescales}
\end{figure}

Figure \ref{fig:timescales} shows $t_\mathrm{coll}(D)$ for the fully populated collisional cascade of the spalled and the recondensed populations. We see that the collisional timescale decreases with decreasing fragment diameter $D$ until a minimum is reached for a size $D^*$ that is greater than the blowout size $D_\mathrm{bl}$. For fragments with $D_\mathrm{bl}<D<D^*$, the collisional timescale increases again because of the absence of $D<D_\mathrm{bl}$ grains, which would have been potentially destructive collisional partners.

However, in the small-size domain, the survival time of grains might be imposed not by $t_\mathrm{coll}$, but by another crucial mechanism: Poynting-Robertson drag (PR-drag), which is also caused by the radiation of the star and makes particles to spiral into the star\footnote{We shall ignore stellar wind drag, an effect similar to PR-drag caused by particles ejected from the star, since its importance depends e.g.\ on the stellar activity.}. The typical timescale for the inward drift is given by  \citep[e.g.][]{Burns_etal_1979}
\begin{equation}\label{eq:t_PR}
t_\mathrm{PR}(D)=\frac{cr^2}{4GM_*\beta(D)}
\end{equation}
We estimate $t_\mathrm{PR}(D)$ for the present set-up and plot it as a dashed green line on Figure \ref{fig:timescales}. As can be seen, this line intersects the $t_\mathrm{coll}(D)$ line at a size $D_\mathrm{PR}\geq D_\mathrm{bl}$. Particles with sizes below $D_\mathrm{PR}$ will thus be removed from the disk before they have had the time to significantly collisionally interact with their environment.  As a consequence, we can expect the size distribution to be strongly depleted in the $D_\mathrm{bl}<D<D_\mathrm{PR}$ domain.

Another important timescale is $t_\mathrm{max}=t_\mathrm{coll}(D_\mathrm{max})$ that sets the timescale for the evolution of the system's total mass, since, as a consequence of the $A=2.5$ power law size distribution, the mass is essentially contained in the largest bodies of the collisional cascade (while most of the cross-section is contained in the smallest bodies). This timescale will thus give the typical survival time of the post-ejection cloud. The system's total mass evolution can be described by \citep[e.g.][]{Wyatt_etal_2007}
\begin{equation}\label{Mtot_vs_t}
M_\mathrm{tot}(t)=\frac{M_\mathrm{tot}(0)}{1+t/t_\mathrm{max}(0)}
\end{equation}
where $M_\mathrm{tot}(0)=M_\mathrm{e}(>v_\mathrm{esc})$ and $t_\mathrm{max}(0)$ is the collisional timescale of the largest fragment at $t=0$.

Finally, for the recondensed population, an important timescale is $t_\mathrm{coll,ini}$, the collisional timescale of the initially monosized population. This timescales sets the duration of the initial phase that transforms the monosized population into a fully populated collisional cascade following the power law of equation \ref{PSD_powerlaw}.

Tables \ref{tab:spalled} and \ref{tab:recondensed} list all these important timescales, computed for the different impact scenarios considered. We assumed $r=1$\,AU and a central star with solar parameters.

\begin{table*}[!hp]
\caption{Parameters characterising the collisional evolution of the spalled population.}
\label{tab:spalled} 
\centering 
\begin{tabular}{c c | c c c c c c} 
\hline\hline 
parameter & & \multicolumn{3}{c}{asteroid} & \multicolumn{3}{c}{comet} \\
& & Earth & Mars & Moon & Earth & Mars & Moon\\
\hline
$\mean{e}$ & - & 0.23 & 0.18 & 0.12 & 0.40 & 0.24 & 0.13\\
$D_\mathrm{max}$& m & 3.6 & 5.7 & 8.6 & 1.5 & 2.2 & 3.3\\
$D_\mathrm{PR}(t=0)$& mm & 10.7 & 1.4 & 0.4 & 4.7 & 1.3 & 0.4\\
$t_\mathrm{max}$ & Myr & 40.2 & 9.6 & 4.4 & 16.9 & 7.0 & 3.7\\
$f_\mathrm{max}$& - & $8.0\times10^{-10}$  & $4.2\times10^{-9}$ & $1.3\times10^{-8}$ & $1.1\times10^{-9}$ & $4.0\times10^{-9}$ & $1.2\times10^{-8}$\\
\hline  
\end{tabular}
\end{table*}

\begin{table*}[!hp]
\caption{Parameters characterising the collisional evolution of the recondensed population. $D_\mathrm{PR}$ is not listed since it varies considerably during the collisional evolution.}
\label{tab:recondensed} 
\centering 
\begin{tabular}{c c | c c c c c c} 
\hline\hline 
parameter & & \multicolumn{3}{c}{asteroid} & \multicolumn{3}{c}{comet} \\
& & Earth & Mars & Moon & Earth & Mars & Moon\\
\hline
$\mean{e}$ & - & 0.42 & 0.26 & 0.15 & 0.39 & 0.25 & 0.14\\
$D_\mathrm{max}$& $\mu$m & 152.4 & 302.7 & 602.6 & 152.9 & 298.2 & 591.0\\
$t_\mathrm{coll,ini}$ & yr & $9.0\times10^4$ & $3.2\times10^4$ & $1.1\times10^4$ & $1.1\times10^5$ & $3.3\times10^4$ & $1.1\times10^4$\\
$t_\mathrm{max}$ & yr & $2.1\times10^3$ & $1.3\times10^3$ & $9.3\times10^2$ & $2.9\times10^3$ & $1.5\times10^3$ & $9.5\times10^2$\\
$f_\mathrm{max}$& - & $1.1\times10^{-6}$  & $2.7\times10^{-6}$ & $6.1\times10^{-6}$ & $8.4\times10^{-7}$ & $2.5\times10^{-6}$ & $5.9\times10^{-6}$\\
\hline  
\end{tabular}
\end{table*}

\subsubsection{Results: Luminosity evolution}\label{coll_evolution_res}
\begin{figure}
\includegraphics[width=1\linewidth]{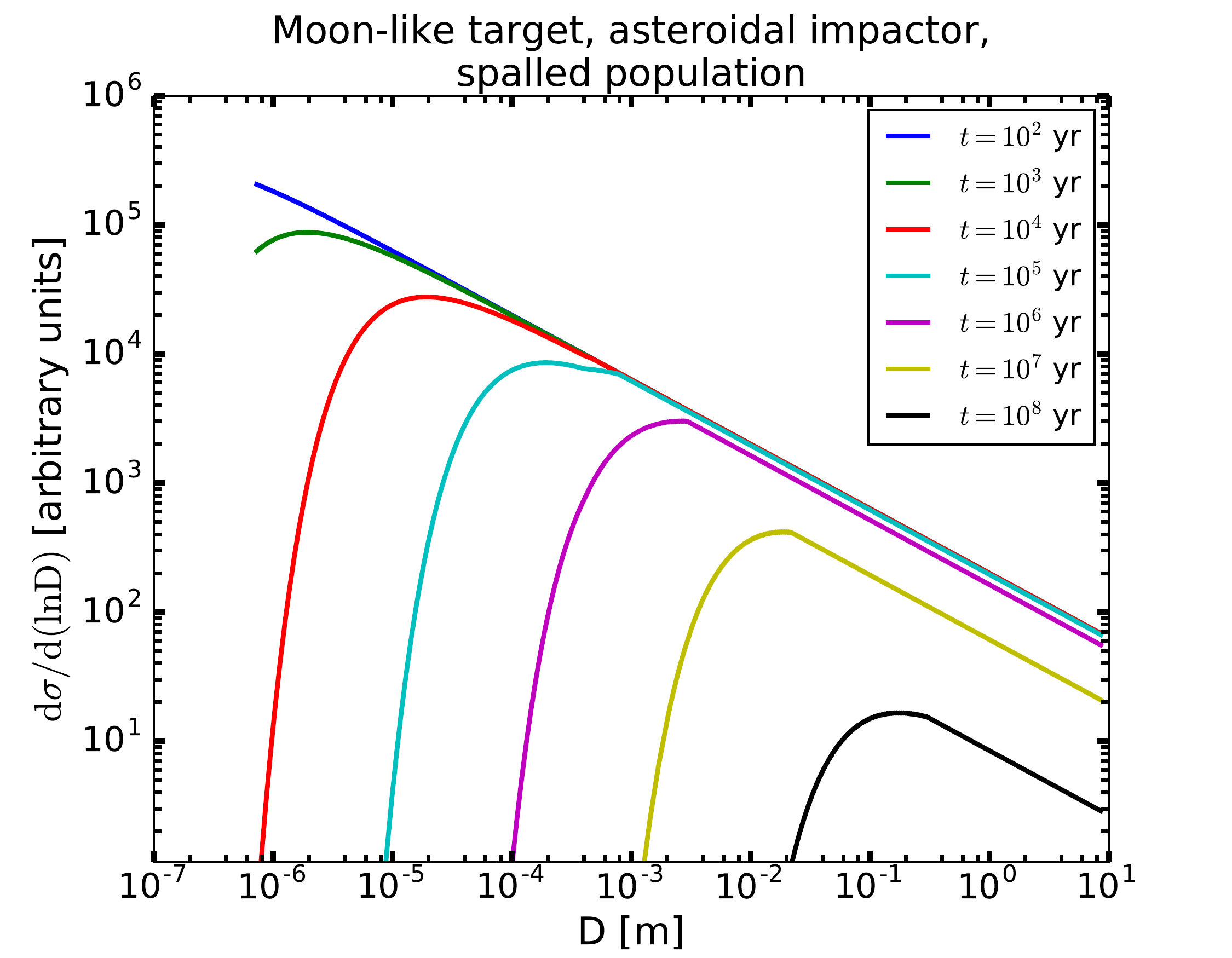}
\includegraphics[width=1\linewidth]{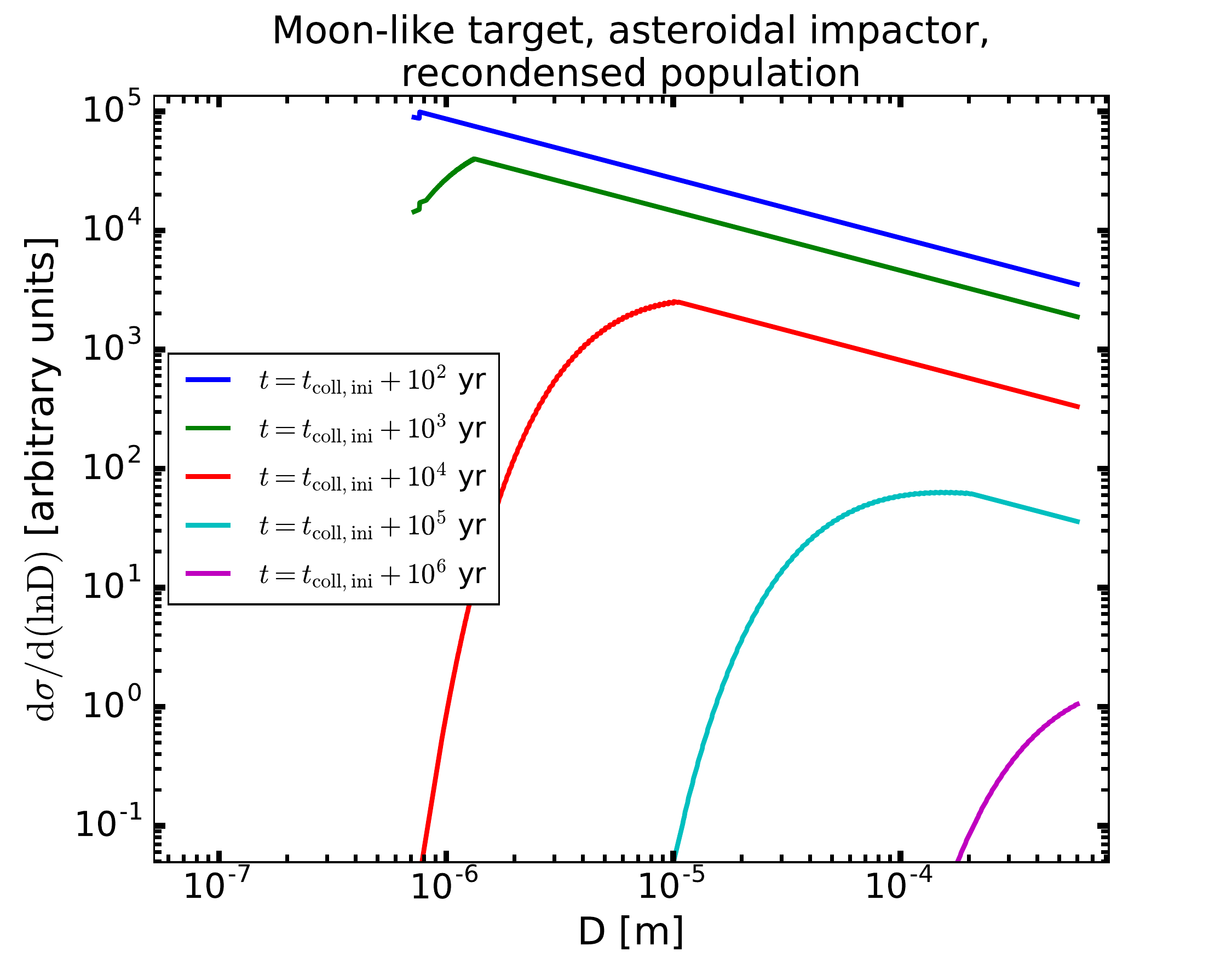}
\caption{Cross section per logarithmic size bin at different epochs if PR drag is taken into account. The chosen units have the advantage of immediately showing what grain sizes contain most of the cross section. We show the spalled population (top) and the recondensed population (bottom, after the initial build-up phase, i.e.\ $t>t_\mathrm{coll,ini}$) in the case of a Moon-like target impacted by an asteroid. For this impact scenario, PR-drag is completely dominating the removal of recondensed grains after $\sim$1\,Myr. For the calculations where PR-drag is ignored, the size distribution always follows a power law down to the blowout size, but is continuously shifted downwards because of the removal of mass by radiation pressure.}
\label{fig:sigma_dist_t}
\end{figure}

Having identified these important timescales, we can schematically compute the collisional evolution of the debris. For simplicity, we assume that no dust other than the newly generated exoplanetary dust is present in the system. Since the competitive effect of collisions and PR-drag will strongly depend on stellar type, we present, for the sake of generality, two limiting scenarios. In the first scenario, the collisional cascade is always fully populated down to the blowout size and the mass decays according to equation \ref{Mtot_vs_t}. In the second scenario, we take PR-drag into account with the following scheme: as for the previous scenario, we assume the mass evolves according to equation \ref{Mtot_vs_t}, but, in addition, we compute at each time step the size $D_\mathrm{PR}$ below which PR-drag significantly suppresses collisions. Grains with a diameter $D<D_\mathrm{PR}$ are then exponentially removed on a timescale $t_\mathrm{PR}(D)$ during the rest of the calculation (i.e.\ smaller grains are removed faster). In other words, we assume that the mass $M_\mathrm{D}$ of grains with $D<D_\mathrm{PR}$ evolves as $\mathrm{d}M_\mathrm{D}/\mathrm{d}t=-M_\mathrm{D}/t_\mathrm{PR}(D).$ Figure \ref{fig:sigma_dist_t} illustrates how the size distribution changes under the influence of collisions and PR-drag. 

We assume that the spalled population evolves as just described from $t=0$ on. However, for the recondensed population, we include an initial build-up phase lasting from $t=0$ to $t=t_\mathrm{coll,ini}$. During this phase, the initially monosized population (all particles have size $D_\mathrm{max}$) is transformed into a fully populated cascade following the power law of equation \ref{PSD_powerlaw} extending down to the blowout size. We shall assume that during this phase, the total cross-section increases linearly\footnote{For a constant total mass, the cross-section increases as the size of the fragments decreases.} and that the mass stays constant. The latter assumption is justified by the fact that the PR-drag timescale of the initial monosized population is longer than $t_\mathrm{coll,ini}$ for all the scenarios we have considered.

We can now estimate how the system's luminosity (in thermal emission) will evolve with time. The thermal emission of the dust can be characterised by the fractional luminosity, defined as the ratio of the infrared (IR) luminosity of the dust and the stellar luminosity:
\begin{equation}
f=\frac{L_\mathrm{IR}}{L_*}
\end{equation}
For blackbody grains (absorption efficiency $Q_\mathrm{abs}=1$), $f=\sigma_\mathrm{tot}/(4\pi r^2)$ with $\sigma_\mathrm{tot}$ the total cross-section of the fragments. Since $\sigma_\mathrm{tot}$ is known at any time, the calculation outlined above directly translates into a curve of fractional luminosity versus time.

\begin{figure}
\includegraphics[width=1\linewidth]{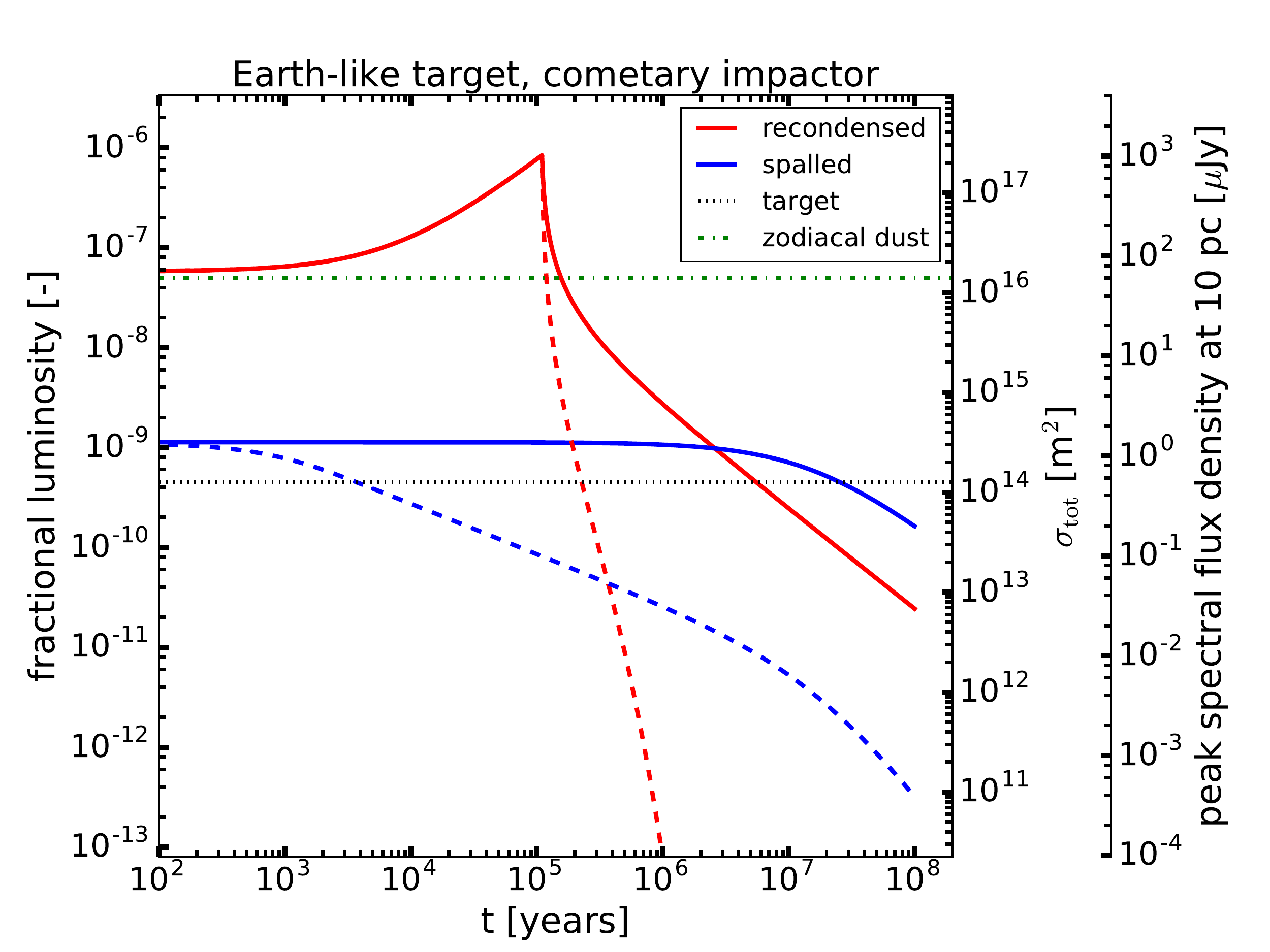}
\includegraphics[width=1\linewidth]{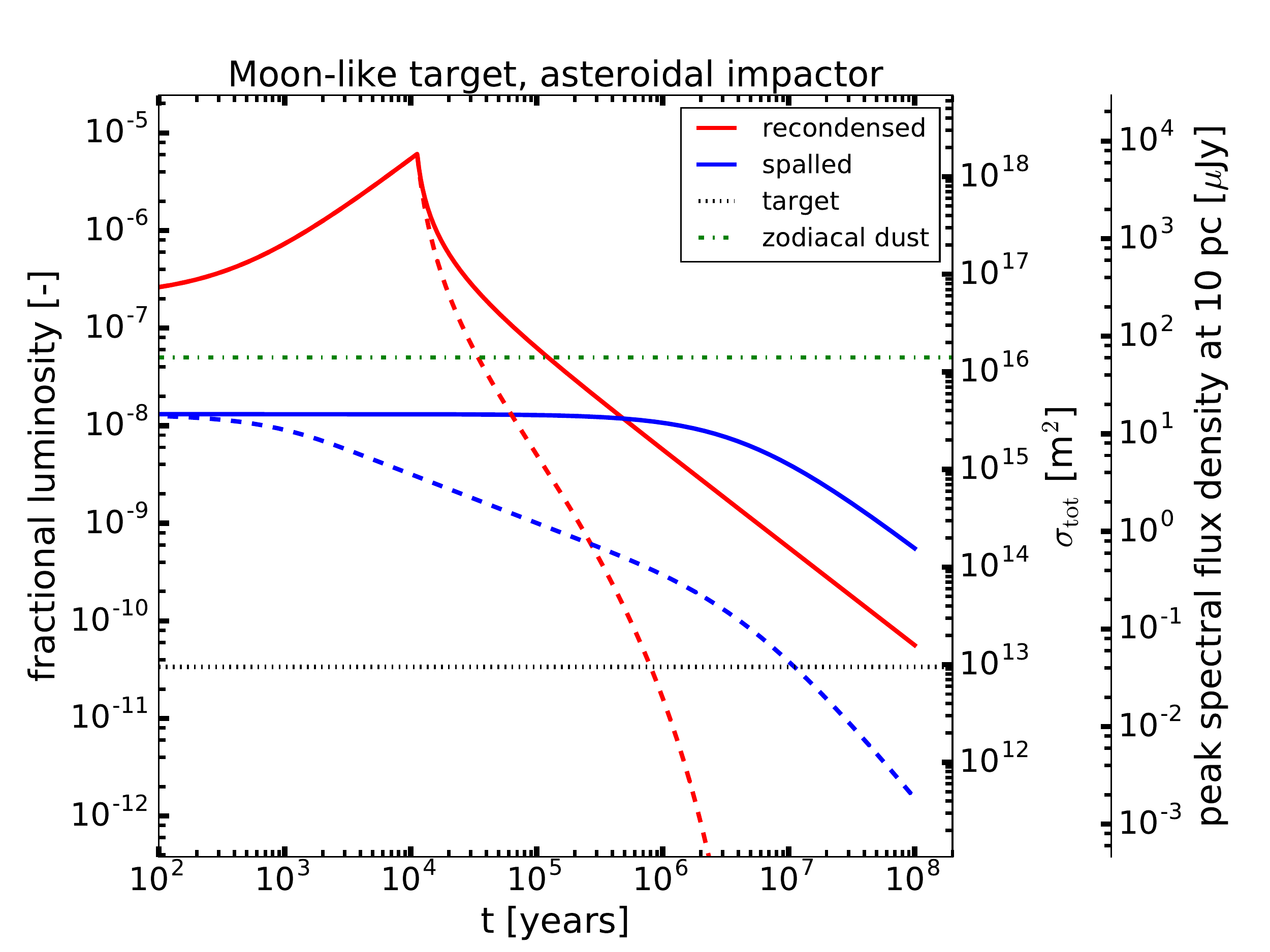}
\caption{Fractional luminosity (left axis), total cross-section and peak spectral flux density from thermal emission (right axis) as a function of time after the impact event for an Earth-like target impacted by a comet (top) and a Moon-like target impacted by an asteroid (bottom). For the peak spectral flux density, we assumed a distance of 10\,pc and that the grains act as black bodies with a temperature of 279\,K (the equilibrium temperature at 1\,AU from a Sun-like star). Solid lines correspond to the calculations where PR-drag is ignored and the cascade is always fully populated down to the blowout size. The dashed lines were computed taking PR-drag into account following the scheme described in the text. The fractional luminosities of the impacted planet and the zodiacal dust are also shown for comparison.}
\label{fig:frac_lum}
\end{figure}

Figure \ref{fig:frac_lum} illustrates the temporal evolution of the fractional luminosity for two different impact scenarios: asteroidal body on a Moon-like target, and cometary body on an Earth-like target. We choose this two scenarios since they correspond to the maximum and minimum escaping mass respectively. Several important results and trends can be seen in figure \ref{fig:frac_lum}. The first one is that the spalled and recondensed populations have significantly different evolutions. The evolution of the recondensed debris can be divided into two phases: 1) an initial luminosity increase phase, as the collisional cascade progressively develops from the initially monosized grains. The increase in $f$ reflects the fact that the total geometrical cross section is increasing as more and more small grains are produced (since, for a $A=2.5$ size distribution, the cross section is dominated by the smaller particles). This first phase stops once collisional steady state is reached, and the start of the steady-state phase corresponds to the peak luminosity reached by this population. After that, the profile of the size distribution no longer evolves (if PR-drag is ignored) and one enters phase 2), where the luminosity progressively decreases as the total mass of the population gradually decreases. The duration of phase 1) is $10^5$ years for the Earth/comet case and $10^4$ years for the Moon/asteroid case. The typical timescale for phase 2) (i.e.\ the timescale for a 50\% luminosity decrease) strongly depends on the importance of PR drag: it can be as long as  $\sim$$10^3$ years without it, but the luminosity decay can be significantly faster with efficient PR drag. We note that this luminosity evolution in two phases bears some similarities to the scenario proposed by \citet{Meng_etal_2014,Meng_etal_2015} to explain the very high and short-lived $f$ values of some ``extreme'' debris disks, which could be due to the vapourisation, mono-size recondensation and subsequent collisional evolution of the debris produced in the aftermath of a very violent collision. The difference is in the scale of the phenomenon, the fractional luminosity of these extreme debris disks reaching values above $10^{-2}$, instead of just $10^{-5}$ here (see below).

For the spalled population, there is no initial luminosity increase, as we have assumed that a size distribution in $N(>D)\propto D^{-2.5}$ holds right from the start. What we observe here is only a progressive luminosity decrease as this population's total mass decreases (note that the log-log scale partially hides the fact that this decrease starts right from $t=0$ and has a rate that actually slows down with time). Here again, PR drag can have a very strong effect, shortening the typical luminosity decrease timescale from a few $10^7$ down to a few times $10^3$ years for both the Earth and Lunar cases.

A crucial result is that the system's total luminosity is dominated, during a long period, by the recondensed population. This is due to two cumulative causes: 1) for all the scenarios considered, there is initially at least four times more mass in the recondensed dust than in the spalled fragments, and 2) the recondensed population is made of much smaller grains, which further increases its cross section and thus its luminosity. The duration of the recondensed-matter-dominated period is of the order of $10^6$ years in the absence of PR drag, and is shortened to a few times $10^5$ years when it is present. This period eventually ends because the collisional removal timescale is shorter for the recondensed population (because of the smaller size of the largest fragments). On long timescales, the luminosity gets dominated by the spalled debris.

For most of its evolution the debris cloud's total fractional luminosity is at a level below $10^{-7}$. The only exception is the initial $\sim$$10^5$ years phase corresponding to the luminosity increase period due to the recondensed population, and the peak value it reaches just when collisional steady-state is reached. This peak value is $\sim$$10^{-6}$ for the Earth/comet case,  and $10^{-5}$ for the Moon/asteroid one, but this initial phase is short lived.  After that, the luminosity continuously decreases. After $1-2\times10^5$ years, it drops below $5\times10^{-8}$, which is approximately the luminosity of the solar system's zodiacal dust (see the green dotted line in figure \ref{fig:frac_lum}). And, for the Earth/comet case, it even eventually drops below the value for the Earth itself (for a more detailed discussion on the luminosity and observability of the debris cloud, see sections \ref{dust_thermal_emission} and \ref{dust_scattered_light}).

\subsection{Oceanic impacts and impacts into ice}\label{subsec:oceanic_impacts}
We shall briefly describe what to expect if the impactor strikes an ocean or an ice crust rather than rock. First of all, we note that Earth's oceans can be considered thin for impactors of the size we study here. For example, using the criterion derived by \citet{Zahnle_1990}, the minimum impactor diameter to reach the floor of a 4\,km deep ocean (typical ocean depth on Earth) is $\sim$3.5\,km for an asteroid (assuming a density of 3000\,kg/m$^3$) and $\sim$6\,km for a comet (1000\,kg/m$^3$). Thus, even in the case of an oceanic impact, one can still expect rocky material to be ejected if the ocean is not too deep. In addition, the water itself can be accelerated to escape velocity, ending up as water gas or water ice in a circumstellar orbit. In order to calculate the maximal amount of escaping water (i.e. assuming a very deep ocean), we adjust the target density $\rho_\mathrm{t}$ and the constants of the \citet{Housen_Holsapple_2011} model used in section \ref{subsec:totalejected}, leaving the other parameters unchanged. This results in escaping water masses comparable to the rock masses in table \ref{tab:impact_results} (typically within $\sim$15\%). If the impactor strikes an icy surface, we only change $\rho_\mathrm{t}$, but leave the \citet{Housen_Holsapple_2011} constants at their value for rock for a first order estimation. In this case, the escaping masses are within $\sim$50\% of the masses in table \ref{tab:impact_results}. Most of the ice will be melted or vaporised. For example, \citet{Pierazzo_Melosh_1999} assumed a pressure of incipient vaporisation of 4.5\,GPa and a pressure of complete vaporisation of 43\,GPa.

In summary, the water mass escaping from an oceanic impact or an impact into ice is comparable to the escaping rock mass from a land impact. However, water gas is expected to be quickly photodissociated. For example, the typical lifetime of a water molecule at a distance of 1\,AU from the Sun is less than a day \citep{Crovisier_1989}, making the probability to detect water gas ejected from an exoplanet vanishingly small. Water gas could still be present if it is continuously produced by sublimation from water ice \citep{Lisse_etal_2012}. But water ice also quickly sublimates unless the ice is very pure \citep{Lisse_etal_2012}. Overall, the prospects to detect water ejected from an exoplanet in the terrestrial zone are not promising given its short lifetime.

\section{Discussion}\label{Sec:Discussion}
\subsection{Uncertainties}
We will now discuss various uncertainties in our modelling and their influence on the derived results, primarily the fractional luminosity.
\subsubsection{Size of the largest fragment}
A crucial parameter for the cloud's evolution is the size of the largest fragments, both for the spalled and the recondensed population, and this for two reasons: 1) because, for the same total mass, changing this maximum size will necessarily change the debris' total luminosity (debris clouds with smaller largest sizes will be more luminous because of their higher geometrical cross section), and 2) because this size will also control the global lifetime of the collisional cascade and thus the long-term evolution of its luminosity.

We first note that the size of the largest fragment in the spalled population is subject to considerable uncertainty. For example, the study of secondary craters on icy satellites showed that the Grady-Kipp distribution (equation \ref{eq:Dmax}) can underestimate the size of the fragments that create the secondary craters by orders of magnitude \citep{Singer_etal_2013}. On the other hand, fragments launched with escape velocity (i.e.\ those ejected with higher velocity than the secondary-crater-forming fragments) are expected to break up in flight, meaning that the Grady-Kipp distribution can still be a reasonable approximation for our purposes.

For a $A=2.5$ size distribution, the cloud's maximum luminosity scales as $D_\mathrm{max}^{-0.5}$, while the collisional timescale for its long-term evolution scales approximately as $D_\mathrm{max}^{0.5}$. To get an idea of the effect the uncertainty on the size of the largest fragment has on our results, we performed additional calculations assuming that the largest fragment is 10 times larger than predicted by equation \ref{eq:Dmax}. This results in fractional luminosities that are lower by not more than a factor of $\sim$3 compared to the values presented in table \ref{tab:spalled}. The collisional timescale of the largest fragment is increased by a factor of 5--6.

There also is evidence that the exponent that describes the dependance of $D_\mathrm{max}$ on the launch velocity $v_\mathrm{e}$ is not constant, but depends on the impactor's size \citep[][and unpublished data]{Singer_etal_2013}. For simplicity, we ignored such a dependence.

The size of the largest fragment in the recondensed population is also uncertain. The equations of \citet{Johnson_Melosh_2014} provide order of magnitude estimates of the mean particle size, while the size dispersion remains unknown. As for the spalled population, changing the size of the largest fragment will both affect the instant luminosity of the fragments and the timescale for their collisional evolution. We find that increasing the size of the largest fragment of the recondensed population by a factor of 10 results in a similar decrease of fractional luminosity and increase of the collisional timescale as for the spalled population.

\subsubsection{Size distribution}
Another uncertainty in our modelling comes from the profile of the fragment size distribution. The profile is of importance because, for a given total mass of debris, it will determine how the geometrical cross section is distributed as a function of particle sizes and what the total value of the geometrical cross section is. For both the spalled and the recondensed population we assumed a power law index of $A=2.5$ for the cumulative size distribution (equation \ref{PSD_powerlaw}), motivated by the work of \citet{Buhl_etal_2014} and the fact that this corresponds to a collisional cascade in steady-state. However, according to \citet[][p.\ 91]{Melosh_1989}, the exponent commonly ranges between 2.4 and 2.7, but may be as small as 1.2 or larger than 3.

To explore the influence of power law index, we consider two limiting values: $A=2$, a shallower distribution for which the \emph{geometrical cross section} is evenly distributed with particle sizes, and $A=3$, a steeper value for which the \emph{mass} is evenly distributed with sizes. For $A = 2$, we find that the fractional luminosity is reduced by a factor $\sim$200 for the spalled population and by a factor 2--4 for the recondensed one. This difference is simply due to the fact that the size range of the distribution is much wider for the spalled population, so that changing $A$ will have a much stronger impact. Conversely, for $A=3$, we find an increase of the fractional luminosity, which reaches $\sim$200 for the spalled population and 3--4 for the recondensed one.

\subsubsection{Radiation forces on small grains}
We assume that grains of any size act as black bodies when calculating the radiation force and the PR-drag timescale (equations \ref{eq:radiation_force}, \ref{eq:beta} and \ref{eq:t_PR}). This is in fact only valid for grains larger than 20\,$\mu$m in the solar system \citep{Gustafson_1994}. However, the general trend that $\beta(D)$ increases as $D$ becomes smaller holds down to micron sized particles. Thus the error introduced by assuming black body grains is probably small compared to the other uncertainties in the model.

\subsubsection{Re-accretion of particles by the impacted exoplanet}
\citet{Jackson_Wyatt_2012} showed that re-accretion of debris onto the impacted planet (or other planets in the system) can be an important removal mechanism. In their simulation, after 10\,Myr, 20\% of the debris created during the Moon-forming impact are re-accreted by Earth, 17\% by Venus and 8\% are ejected from the solar system through scattering by Jupiter. The amount of re-accreted material actually changes with the number of planets in the system. For example, additional planets can reduce the re-accretion rate by stirring the debris. Using the analytical expressions given in \citet{Genda_etal_2015} and \citet{Jackson_Wyatt_2012} with our parameters, the depletion timescale of fragments by accretion is tens (Earth-sized target) to hundreds (for Moon-sized target) of millions of years, suggesting re-accretion is a minor issue, especially for small exoplanets. However, these analytical estimates do not account for the initial asymmetry of the debris disk which can enhance the accretion rate significantly \citep{Jackson_Wyatt_2012}.

\subsubsection{Mixing of target and projectile material}
We are primarily interested in the dust originating from the impacted exoplanet, but the ejecta contain material from the impactor as well, potentially resulting in a pollution of the signal \citep{Morlok_etal_2014}. The problem might be more pronounced for impacts onto large planets where the escaping mass can be comparable to the impactor mass (see table \ref{tab:impact_results}). Escaping ejecta from the Chicxulub impact can be dominated by impactor material depending on the impact parameters \citep{Artemieva_Morgan_2009}. On the other hand, escaping ejecta from the Moon \citep{Artemieva_Shuvalov_2008} or Titan \citep{Korycansky_Zahnle_2011} were estimated to often (but not always) be dominated by target material. Judging from values for the mean pressure experienced by most parts of the impactor \citep{Melosh_1989}, we note that the escaping impactor material is melted or vaporised to a large degree and does not contribute to the spalled population significantly. Thus, at least for the spalled population, pollution by impactor material should not be a problem.

\subsection{Spectral type of the host star}
Throughout our modelling in section \ref{Sec:Modelling}, we assumed a Sun-like host star. However, statistical analysis of exoplanetary transit data suggests that M-dwarfs should commonly have planets in their habitable zone \citep[e.g.][and references therein]{Dressing_Charbonneau_2015}. We note that several aspects of our modelling would need to be revised when considering such systems. Because of the increased orbital velocity of exoplanets in the habitable zone of M-dwarfs, the impact event would be more violent and damaging to the ejecta, but also result in larger escaping masses. The collisional evolution of the debris would proceed significantly faster (see equation \ref{eq:t_coll}). Also, radiation pressure from M-dwarfs is usually too weak to remove any grains \citep[e.g.][]{Matthews_etal_2007}. Instead, stellar wind forces can become important around this type of stars \citep[e.g.][]{Plavchan_etal_2005}.

\subsection{Is the presence of impact generated dust detectable?}
We will now discuss whether the presence of dust generated by an impact event of the type considered in section \ref{Sec:Modelling} can be detected by present or future instrumentation. We consider both thermal emission and scattered light. In section \ref{subsec:dust_composition}, we will explore the possibility to study the dust composition.

\subsubsection{Detection of the dust thermal emission}\label{dust_thermal_emission}
The thermal emission of a dust grain with diameter $D$ at a distance $r$ from the star depends on the grain's temperature $T$, which can be determined (numerically) from the equation balancing absorption and emission:
\begin{multline}
\int\frac{L_\nu}{4\pi r^2}\left(\frac{D}{2}\right)^2\pi Q_\mathrm{\nu,abs}(D)\mathrm{d}\nu\\
 = \int\pi B_\nu(T)4\pi\left(\frac{D}{2}\right)^2Q_\mathrm{\nu,em}(D)\mathrm{d}\nu
\end{multline}
where $\nu$ is the frequency, $B_\nu(T)$ the Planck function and $L_\nu$ is the specific luminosity (i.e.\ the luminosity per frequency interval) of the star. If the star is modelled as a black body of effective temperature $T_*$, then $L_\nu=4\pi R_*^2\pi B_\nu(T_*)$ with $R_*$ the stellar radius. $Q_\mathrm{\nu,abs}$ is the absorption efficiency defined as the ratio between absorption cross-section and geometrical cross-section and depends on grain size, material and shape. $Q_\mathrm{\nu,em}$ is the emission efficiency, the ratio between the surface area a blackbody would need to emit the same flux divided by the actual surface area of the grain. Because of Kirchhoff's law, $Q_\mathrm{\nu,abs}=Q_\mathrm{\nu,em}$. Thus, once the temperature of the grain is known, the emitted spectral flux density observed at a distance $d$ is written
\begin{equation}
F_\nu=\pi B_\nu(T)\cdot4\pi\left(\frac{D}{2}\right)^2Q_\mathrm{\nu,abs}(D)\cdot\frac{1}{4\pi d^2}
\end{equation}
where $d$ is the distance of the observer. The position and strength of spectral features is thus encoded in $Q_\mathrm{\nu,abs}$, which can be calculated by using data obtained from laboratory experiments.

The presence of circumstellar dust can for example be inferred by an infrared (IR) excess above the stellar photosphere, i.e.\ IR photometry yields a higher flux than what would be expected from the star alone due to the additional thermal emission from the dust grains. The sensitivity of a certain instrument can then be characterised by the smallest fractional luminosity $f_\mathrm{min}$ it is able to detect. Actually, $f_\mathrm{min}$ depends on the spectral type of the star, the wavelength of observation and the temperature of the dust (or equivalently its radial distance from the star). For dust in the terrestrial region around a solar-type star, the \textit{Spitzer Space Telescope} was sensitive to fractional luminosities of the order of a few times $10^{-5}$ to $10^{-4}$ \citep{Wyatt_2008,Roberge_etal_2012}. Our calculations show that the fractional luminosity from dust originating in an impact event only comes somewhat close to this value for the recondensed population in the case of a Moon-sized target. Fractional luminosities from the spalled population are at least three orders of magnitude smaller. Unless impacts occur at high frequency such that dust can accumulate (e.g.\ in a Late Heavy Bombardment scenario), a \textit{Spitzer}-like telescope is certainly unable to detect dust from the spalled population, and probably even form the recondensed population. Nulling interferometers can achieve better sensitivities. These instruments combine beams from different telescopes such that light from the star interferes destructively, effectively blocking ("nulling") the star and leaving only light from its surrounding, e.g.\ from circumstellar dust or exoplanets. The ground-based Large Binocular Telescope Interferometer (LBTI) has recently started operating and is expected to have a sensitivity equivalent to a few times the level of the zodiacal dust \citep{Roberge_etal_2012,Weinberger_etal_2015}. The zodiacal dust's fractional luminosity is estimated to be $f_\mathrm{zodi}=10^{-8}-10^{-7}$ \citep{Dermott_etal_2002,Nesvorny_etal_2010}, which is in-between the fractional luminosities achieved by the recondensed and the spalled population respectively (figure \ref{fig:frac_lum}). Thus, LBTI has good prospects to detect dust generated in impact events as discussed in the present work.

For the direct detection of Earth-analogs, nulling interferometers operating in space such as \textit{Darwin} or TPF-I have been proposed \citep[e.g.][]{Cockell_etal_2009}. Such instruments would have point source sensitivities on the order of a microjansky or smaller. Assuming the dust grains radiate as black bodies and a distance to the target of 10\,pc, we find the maximum dust thermal emission of the spalled population to be larger than 0.5\,$\mu$Jy for all the impact scenarios considered in this work. We conclude that future space nulling interferometers are able to detect and image dust from impacts, even when considering the fact that the dust is not a point source \citep{Rottgering_etal_2003}. How long after the impact the signal remains detectable depends on the total amount of dust and how strongly PR-drag is influencing the dust dynamics. Considering figure \ref{fig:frac_lum} and taking $\sim$1\,$\mu$Jy as a limit for detectability, for an Earth-sized planet, the spalled population becomes undetectable very quickly (within $\sim$$10^3$ years) under the influence of PR-drag, while the recondensed population remains detectable for $\sim$$10^5$ years. If PR-drag is ignored, the recondensed population remains detectable for nearly $10^7$ years, and the spalled population even longer. In the more favourable case of a smaller, Moon-sized planet, PR-drag has made the spalled population undetectable after $\sim$$10^5$ years, the recondensed population remaining detectable only slightly longer. If PR-drag is ignored, the spalled population remains detectable for longer time ($\sim$$10^8$ years) than the recondensed population ($\sim$$10^7$ years).

\subsubsection{Detection of scattered light}\label{dust_scattered_light}
Apart from absorbing and re-emitting in the thermal infrared, dust grains can also scatter light coming from the host star. The star is in general outshining the scattered light from the dust disk by many orders of magnitude. This difficulty can be overcome by using a coronagraph, a device that blocks direct light from the star. Such an instrument can be characterised by the achievable contrast, i.e.\ the minimum brightness of an object in order to be detectable divided by the stellar brightness, and the inner working angle, i.e.\ the smallest distance from the star at which an object is detectable.

The spectral flux density scattered into a solid angle $\mathrm{d}\Omega$ about the direction $\Omega$ by a particle of diameter $D$ located at distance $r$ from the star can be written
\begin{multline}
\mathrm{d}F_\mathrm{\nu,sca}=\\
\frac{1}{4\pi d^2}\cdot\left(\frac{R_*}{r}\right)^2 F_{\nu,*} \cdot\left(\frac{D}{2}\right)^2\pi Q_\mathrm{\nu,sca}(D)\cdot\varphi_\nu(\Omega)\mathrm{d}\Omega
\end{multline}
where $F_{\nu,*}$ is the spectral flux density at the stellar surface, $R_*$ is the stellar radius, $Q_\mathrm{\nu,sca}(D)$ is the scattering efficiency (i.e.\ the scattering cross-section divided by the geometric cross-section) and $\varphi_\nu(\Omega)$ is the phase function describing the directional dependance of the scattering process.

To estimate the amount of scattered light from the ejected dust, we assume isotropic scattering (i.e.\ $\varphi_\nu(\Omega)=\mathrm{const.}=1/4\pi$) and model the stellar emission as a black body: $F_{\nu,*}=\pi B_\nu(T_*)$ with $T_*$ the effective temperature of the Sun. We also assume $Q_\mathrm{\nu,sca}(D)=1$, which is appropriate for black body grains large compared to the light's wavelength\footnote{This follows from the fact that $Q_\mathrm{abs}+Q_\mathrm{sca}=2$ for large grains \citep[e.g.][]{Bohren_Huffman_1998}.}, which is a good approximation since the star's emission peaks in the visible wavelength region. With this assumptions, the contrast is actually the same as the fractional luminosity we calculated earlier. We now consider whether space-based coronagraphs that might become available in the future can detect the scattered light from the dust disk. Such instruments include the \textit{Terrestrial Planet Finder Coronagraph} \citep[TPF-C,][]{Levine_etal_2006} or the more recently studied \textit{Exo-C} \citep{Stapelfeldt_etal_2015} and \textit{Exo-S} \citep{Seager_etal_2015}. The latter would involve a "starshade" tens of metres in size flying ten-thousands of kilometres from the telescope, instead of an internal coronagraph. Since the primary purpose of such instruments would be imaging of Earth-twins (i.e.\ Earth-like planets in the habitable zone of a Sun-like star), their sensitivity is specified in terms of the contrast an exoplanet (point source) needs in order to be detectable. A typical value is $10^{-10}$, the contrast of Earth. However, we are interested in detecting an extended object. Thus, we calculate the contrast of the dust belt \emph{per resolution element} (assuming a resolution of $1.22\lambda/D$ with $D$ the telescope diameter), assuming the belt has the previously calculated width $\Delta r$ and is observed from a distance of 10\,pc. We also take into account that those parts of the belt at an angular separation from the star smaller than $4\lambda/D$ are blocked by the coronagraphic mask \citep {Levine_etal_2006}. The contrast per resolution element can be compared to the aforementioned minimum point-source contrast detectable by the instrument. For TPF-C with an 8\,m mirror, the dust belt emission typically covers 2 (for an edge-on belt mostly blocked by the coronagraph) up to $\sim$20 (for a face-on belt) resolution elements. To be detectable by TPF-C, its total contrast (or fractional luminosity) should therefore be around $10^{-9}$. Comparing with figure \ref{fig:frac_lum}, such a contrast is achieved by the recondensed population for more than $10^5$\,yr even under the influence of PR-drag. The spalled population can also be detected in scattered light in the case of smaller exoplanets. For example, in the Moon/asteroid case, such a contrast is achieved for a few times $\sim$$10^4$\,yr (PR-drag) up to $\sim$$10^8$\,yr (no PR-drag). We note that a face-on disk would probably be more difficult to detect, because specialised data reduction processes can make use of the asymmetry of the target (exoplanet or edge-on disk) to optimise sensitivity; this is not possible for a symmetric face-on disk.

\subsection{Is it possible to constrain the dust composition?}\label{subsec:dust_composition}
We have demonstrated that the presence of impact-generated dust can be detected by future, and possibly even current instruments. We go now one step further and ask whether the dust \emph{composition} can be constrained.
\subsubsection{Parametrising the detectability of a generic signal}
To address the possibility of constraining the dust composition, we set up a simple model to explore the detectability of a generic (spectral) signal that is mixed with some background. We assume that the data consist of a total signal $F$ that is the sum of a background component $B$ and the signature of interest $S$:
\begin{equation}\label{fitting model}
F(\lambda)=p_1B(\lambda) + p_2S(\lambda)
\end{equation}
Here $B$ and $S$ are \emph{models} of the background and the signal of interest respectively and are both assumed to be a function of the wavelength $\lambda$, while $p_1$ and $p_2$ are scaling factors. The background component could for example be the component of the impact-generated dust we are not interested in, or it could be exo-zodiacal dust present in the system. We assume that the measurement process adds Gaussian noise with standard deviation $\sigma_\mathrm{i}$ to each measurement point $F_\mathrm{i}$. To what precision can $p_1$ and $p_2$ be determined if one considers them as free parameters and fits them to the noisy data? This question can readily be answered by calculating and inverting the Fisher matrix \citep[e.g.][]{Andrae_2010} of the model described by equation \ref{fitting model}. We assume the instrument measures at $N$ distinct wavelengths $\lambda_\mathrm{i}$ and define a vector $\vec{s}$ by $s_\mathrm{i}=S_\mathrm{i}/\sigma_\mathrm{i}$. This leads to the following expression for the uncertainty on the fitted parameter $p_2$:
\begin{equation}\label{absolute p2 uncertainty}
\sigma_\mathrm{p_2} = \frac{1}{\|\vec{s}\|\: |\sin \phi|}
\end{equation}
where $\|\vec{s}\|$ denotes the Euclidian norm of $\vec{s}$ and $\phi$ is the angle between $\vec{b}$ and $\vec{s}$ defined from the usual scalar product: $\vec{b}\cdot\vec{s}=\|\vec{b}\|\:\|\vec{s}\|\cos\phi$ (with $\vec{b}$ defined analogue to $\vec{s}$). Equation \ref{absolute p2 uncertainty} can be rewritten as a ``signal-to-noise ratio'' (SNR) for the signature we are interested in:
\begin{equation}\label{SNR_general}
\mathrm{SNR}=\frac{p_2}{\sigma_\mathrm{p_2}}=\|\vec{s}_\mathrm{t}\|  \:|\sin\phi|
\end{equation}
where the elements of $\vec{s}_\mathrm{t}=p_2\vec{s}$ correspond to the true signature divided by the measurement uncertainty. If the measurement error does not vary with wavelength ($\sigma_\mathrm{i}=\sigma$, $\forall \mathrm{i}$), the expression can be written
\begin{equation}\label{SNR_const_sigma}
\mathrm{SNR}=\frac{\|\vec{S}_\mathrm{t}\|}{\sigma} \:|\sin\phi|
\end{equation}
Also, in this case, $\phi$ can be computed directly from $\vec{B}$ and $\vec{S}$. Thus, assuming a signal strength $\vec{S_\mathrm{t}}$ for a particular impact scenario, and given the angle $\phi$ characterising the the relation between signature and background, equation \ref{SNR_const_sigma} allows for computation of the measurement error $\sigma$, and in turn the instrument sensitivity needed to detect the signature with a given SNR. We note the following points:
\begin{itemize}
\item The term $|\sin\phi|$ describes the effect of the background. It would disappear if one considers a single parameter model without background.
\item As expected, the signature is easiest to detect when $\vec{b}$ and $\vec{s}$ are orthogonal. Conversely, if $\phi=0$, the model is completely degenerate since background and biosignature are proportional to each other.
\item Both $\|\vec{S}_\mathrm{t}\|$ and $\phi$ depend on the number of data points $N$. However, with denser and denser sampling, $\phi$ tends towards an asymptotic value.
\end{itemize}

In general, the model described in this section is useful to parametrise the detectability of a given signature in a simple way. However, we note it is also an optimistic model since we assume that the components that make up the total signal are exactly known a priori (equation \ref{fitting model}), with only their scalinkg unknown. This is most often not true in practice. For example, the emission spectrum of minerals depends on various parameters such as grain size distribution, grain shape, crystalline form etc.

\subsubsection{Conversion of instrument sensitivities to measurement errors}
The sensitivity specified for a given instrument often applies to the background-limited case\footnote{This means that the noise is dominated by background emission such a zodiacal light. The background emission should not be confused with the background signal discussed in the previous section. The latter is a part of the data where the background emission has been subtracted.}. In the following sections, we will consider observations in the infrared where the dominant contribution to the diffuse infrared night sky brightness, and thus to the photon noise, is the zodiacal light \citep{Leinert_etal_1998}. Thus, we convert instrument sensitivities from the literature into measurement errors ($\sigma$ in equation \ref{SNR_const_sigma}) in the following way. First, we calculate the flux from the zodiacal dust picked up by the telescope beam, using the surface brightness of the zodiacal dust given by \citet{Leinert_etal_1998}. We also compute the "target flux", consisting of flux from the impact generated dust cloud, and the stellar photosphere (unless the telescope can resolve the dust belt such that the star is not contributing)\footnote{We shall ignore additional contribution to the target flux (and thus the photon noise) from possibly present exozodiacal dust in the system, as we find it to be negligible as long as the star contributes (i.e.\ the belt is not resolved). If the belt can be resolved (e.g.\ by TPF-I), the contribution would depend on the amount of exo-zodiacal dust as well as its location in the system (i.e.\ whether it can be spatially separated from the impact-generated dust).}. The measurement error $\sigma$ is then calculated assuming photon-noise (i.e.\ Poisson statistics) in the following way:
\begin{equation}
\sigma=\xi\sqrt{\frac{F_\mathrm{tg}+F_\mathrm{zodi}}{F_\mathrm{zodi}}}
\end{equation}
with $\xi$ the instrument sensitivity from the literature, $F_\mathrm{tg}$ the target flux and $F_\mathrm{zodi}$ the flux from zodiacal dust. This means that we always use a sensitivity value that is increased compared to the literature value. We shall also assume that $\sigma\propto t_\mathrm{exp}^{-1/2}(\Delta\lambda)^{-1/2}$ where $t_\mathrm{exp}$ is the exposure time and $\Delta\lambda$ the spectral resolution.


\subsubsection{Example: thermal emission from calcite}\label{calcite_astrosili}
Calcite is the stable mineral form of calcium carbonate CaCO$_3$. On Earth, calcite is commonly found in sedimentary rocks, particularly in limestone, which is largely formed from marine skeletal fragments and shells. Calcite can also form abiotically through chemical precipitation. It is interesting that the internal crystal structures of biogenic and abiotic calcite differ from each other. The biogenic crystal lattice has anisotropic lattice distortions with possible substitution of COO$^-$ for some of the carbonate groups \citep{Pokroy_etal_2006}. \citet{Berg_etal_2014} found that the presence of organic matter makes biogenic calcite spectroscopically distinguishable from abiotic calcite. 

In an astronomical context, calcite has been detected around both evolved stars \citep{Kemper_etal_2002} and protostars \citep{Ceccarelli_etal_2002,Chiavassa_etal_2005}. The calcite detected around stellar environments has been suggested to form under dry conditions (without the presence of water) through reactions of amorphous silicates and CO$_2$ \citep{Gillot_etal_2009}. The significance of calcite as a biosignature is therefore limited, however, its detection from an impact on an exoplanet would still be interesting. It would provide useful information on the geology of the planetary body. For example, the production of calcite on Earth is dependent on aqueous solution and even if dry calcite formation is possible, wet calcite formation is thermodynamically and kinetically much more favourable. Therefore, the detection of calcite would likely indicate the presence of liquid water at some point in the history of the planet. The presence of calcite reveals the thermal and alkalinity thresholds of the planet (temperature and pH) as well, which are important factors that affect its habitability \citep{Rodriguez-Navarro_etal_2009}. Also, calcite present on an exoplanet may provide a stable, long-term carbon source for microorganisms \citep{Jacobson_Wu_2009}.

A potential problem arises from space weathering, i.e.\ damage induced by, for example, cosmic rays or the solar wind. These effects might change the spectral properties of ejected calcite or even destroy it over extended periods of time. The lack of detailed studies makes it difficult to judge whether this is an important process, and if so, on which timescales it occurs. However, the detection of calcite in young and old circumstellar environments \citep{Kemper_etal_2002,Ceccarelli_etal_2002, Chiavassa_etal_2005} or in meteorites \citep[e.g.][and references therein]{Lee_etal_2014} suggests that calcite can survive in space at least for some time. In addition, we note that if space weathering is primarily affecting the \emph{surface} of the fragments, the problem might be mitigated by the fact that the fragments continuously collide, which regularly exposes fresh surfaces.

We consider the possibility to detect calcite ejected as part of the spalled population (i.e.\ $P_\mathrm{max}<50$\,GPa). The amount of ejected calcite will strongly depend on the geology of the impact site. For the sake of argument, we consider two cases: an upper limit case where 100\% of the spalled ejecta consist of calcite, and a case with a calcite fraction of 2\%, which is the an average volume percentage of limestone in Earth's crust \citep{Marshall_Fairbridge_1999}.


Calcite has its strongest emission features in the far-IR. We investigate whether proposed future far-IR telescopes would be able to detect calcite signatures. The \textit{Space Infrared Telescope for Cosmology and Astrophysics} (SPICA) is a proposed joint Japanese-European mission \citep[e.g.][]{Nakagawa_etal_2014} with a far-IR instrument called SAFARI \citep[e.g.][]{Roelfsema_etal_2014}. We also consider the more ambitious \textit{Single Aperture Far-Infrared Observatory} \citep[SAFIR,][]{Benford_etal_2004,Lester_etal_2004}.

\begin{figure}
\includegraphics[width=1\linewidth]{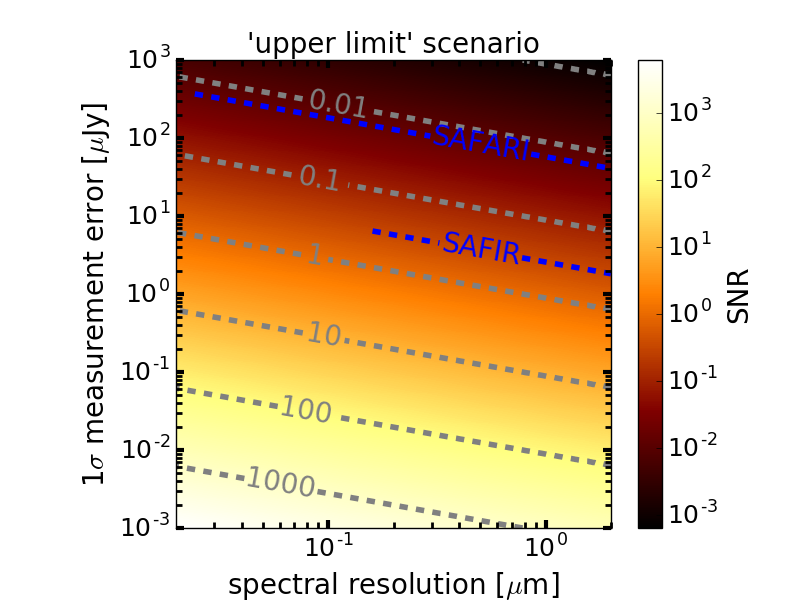}
\includegraphics[width=1\linewidth]{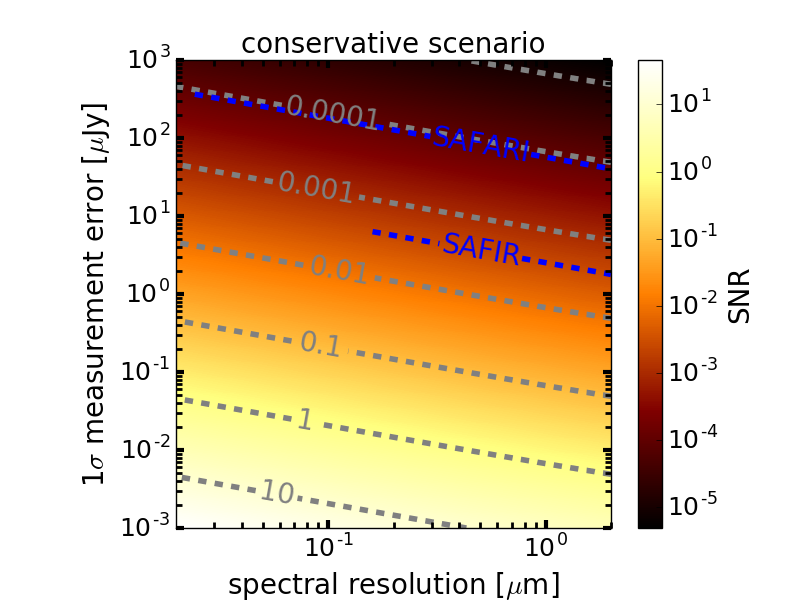}
\caption{The SNR when observing calcite (ejected by an asteroidal impact onto a Moon-like planet) in the mid/far-IR (34 to 150\,$\mu$m) with an astrosilicate background at a distance of 10\,pc. The SNR is calculated from equation \ref{SNR_const_sigma}, for the `upper limit' and the conservative scenario described in the text. The error--resolution lines for two far-IR instruments are also shown, assuming an exposure time of one hour.}
\label{fig:SNR_calcite}
\end{figure}

To calculate the calcite emission spectrum, we compute absorption efficiencies using Mie theory (i.e.\ assuming compact, spherical grains). We use the \texttt{BHMIE} code \citep{Bohren_Huffman_1998_AppendixA} in combination with \texttt{f2py} \citep{Peterson_2009}. The optical constants are taken from \citet[][Jena Database of Optical Constants for Cosmic Dust]{Posch_etal_2007} and extrapolated into the visible wavelength range with a power law. The calculated absorption efficiencies allow us to determine the temperature of the grains and subsequently the thermal emission spectra assuming a distance of 1\,AU from the host star (section \ref{dust_thermal_emission}). For simplicity, we model the stellar spectrum as a blackbody with a temperature equal to the effective temperature of the Sun. Using the calculated emission spectrum, we determine the SNR using equation \ref{SNR_const_sigma}, assuming a specific background, namely astrosilicates \citep{Draine_Lee_1984}, which have been used to model emission from dust in a variety of astronomical objects, in particular debris disk spectra as well as the zodiacal dust. We use the method described above to compute astrosilicate emission spectra, making use of optical constants from \citet{Draine_Lee_1984}. We then consider two limiting scenarios (Moon-like target, asteroidal impactor): a conservative case where we determine the flux from the PR-affected spalled population after $t=10^5$\,yr, with a calcite fraction of 2\%, and an `upper limit' case with a fully populated cascade at $t=0$ and a calcite fraction of 100\%. Figure \ref{fig:SNR_calcite} shows what measurement error and spectral resolution would be necessary to detect calcite. We also plot the error-resolution lines for 1-hour exposures with SAFARI (B.\ Sibthorpe 2015, private communication) and SAFIR \citep{Lester_etal_2005}. The figure shows that SAFARI would not be able to detect calcite even in the optimistic scenario. This is not surprising given that we find a peak calcite flux of only $\sim$0.3\,$\mu$Jy in the `upper limit' scenario, roughly two orders of magnitude smaller than the typical SAFARI sensitivity. On the other hand, a SAFIR-like telescope might be able to detect calcite under favourable circumstances and with long exposure times.

\subsubsection{Example: thermal emission from glassy silica}\label{sec:silica_emission}
The detection of a substantial amount of glassy silica (SiO$_2$) in the debris belt would suggest that the observed dust originates indeed from a violent event that transformed silicates to silica \citep[e.g.][]{Lisse_etal_2012}. This would be in contrast to e.g.\ dust from comet sublimation, since comets show, if at all, only small amounts of silica. Production of glassy silica is expected when silicaceous material is heated to high temperature and then quickly quenched, preventing the formation of a crystalline structure and thus resulting in amorphous material. This can for example occur during an impact event or in a giant collision between planetesimals. A few debris disks, for example those around HD172555 \citep{Lisse_etal_2009} or HD15407\,A \citep{Fujiwara_etal_2012}, show unusually silica-rich dust. As a dust origin, a giant hypervelocity ($>$10\,km\,s$^{-1}$) impact between large rocky planetesimals has been suggested. The events creating the silica-rich dust may have been similar to the Moon-forming event in the solar system. The disk around $\eta$~Corvi also contains abundant amounts of silica, although silicates such as olivine and pyroxene are also present, arguing for a less violent impact event (5-10\,km\,s$^{-1}$) with incomplete silicate-to-silica transformation \citep{Lisse_etal_2012}.

\begin{figure}
\includegraphics[width=1\linewidth]{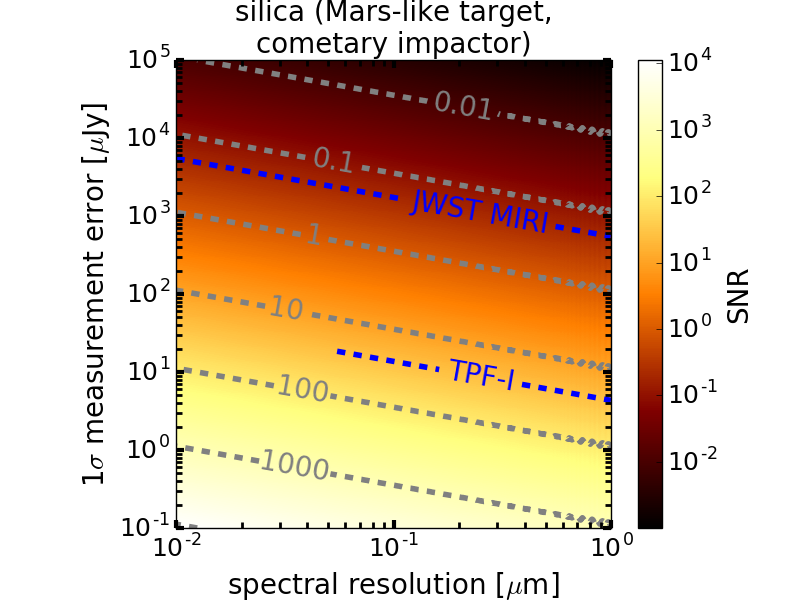}
\caption{The SNR when observing silica (ejected by a cometary impactor onto an Mars-like planet) in the mid-IR (5 to 25\,$\mu$m) with an astrosilicate background at a distance of 10\,pc when the recondensed population is at its peak luminosity. The SNR is calculated from equation \ref{SNR_const_sigma}. Error-resolution lines for two mid-IR instruments are also shown, assuming an exposure time of one hour.}
\label{fig:SNR_silica}
\end{figure}

\begin{figure}
\includegraphics[width=1\linewidth]{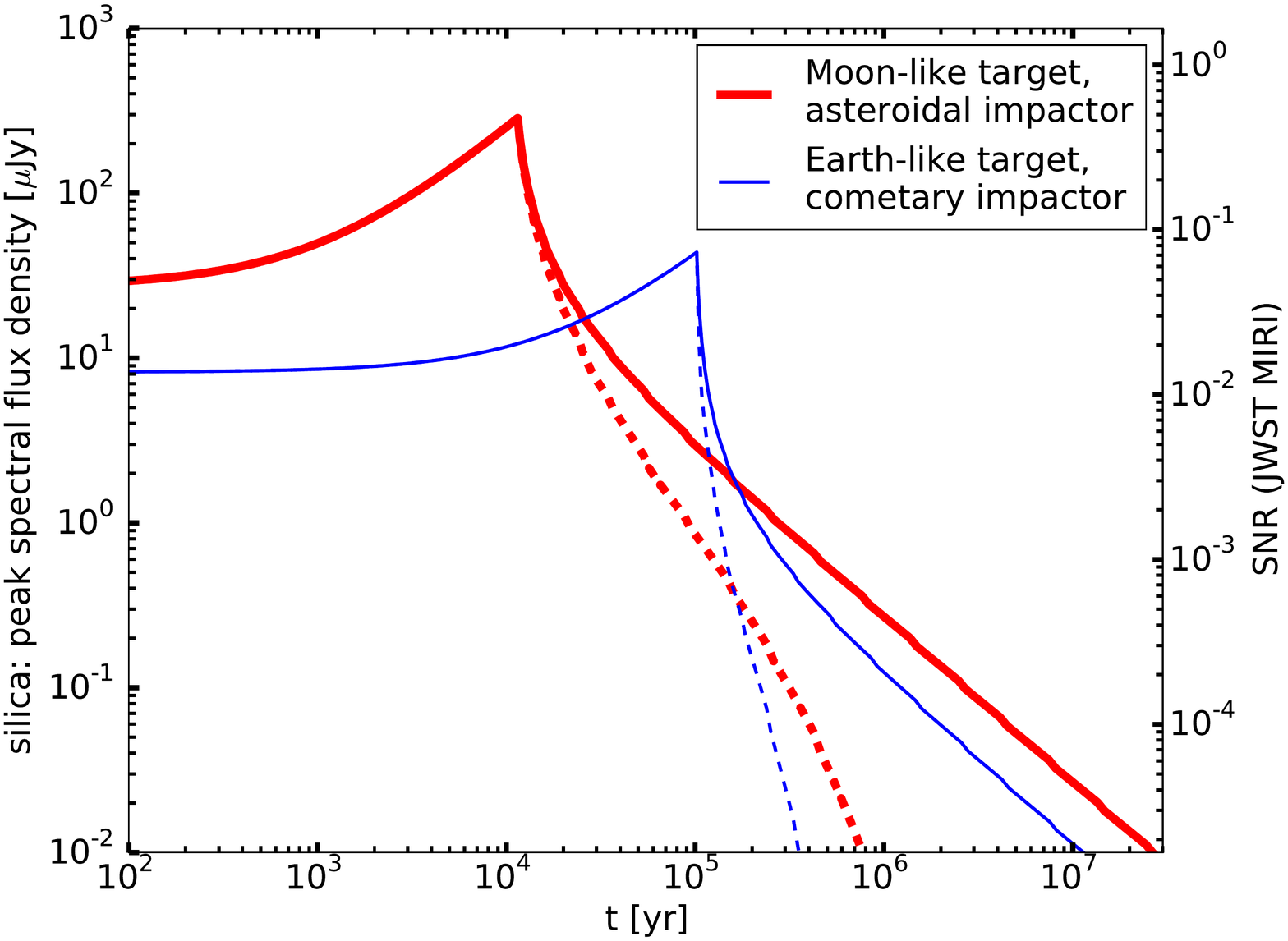}
\includegraphics[width=1\linewidth]{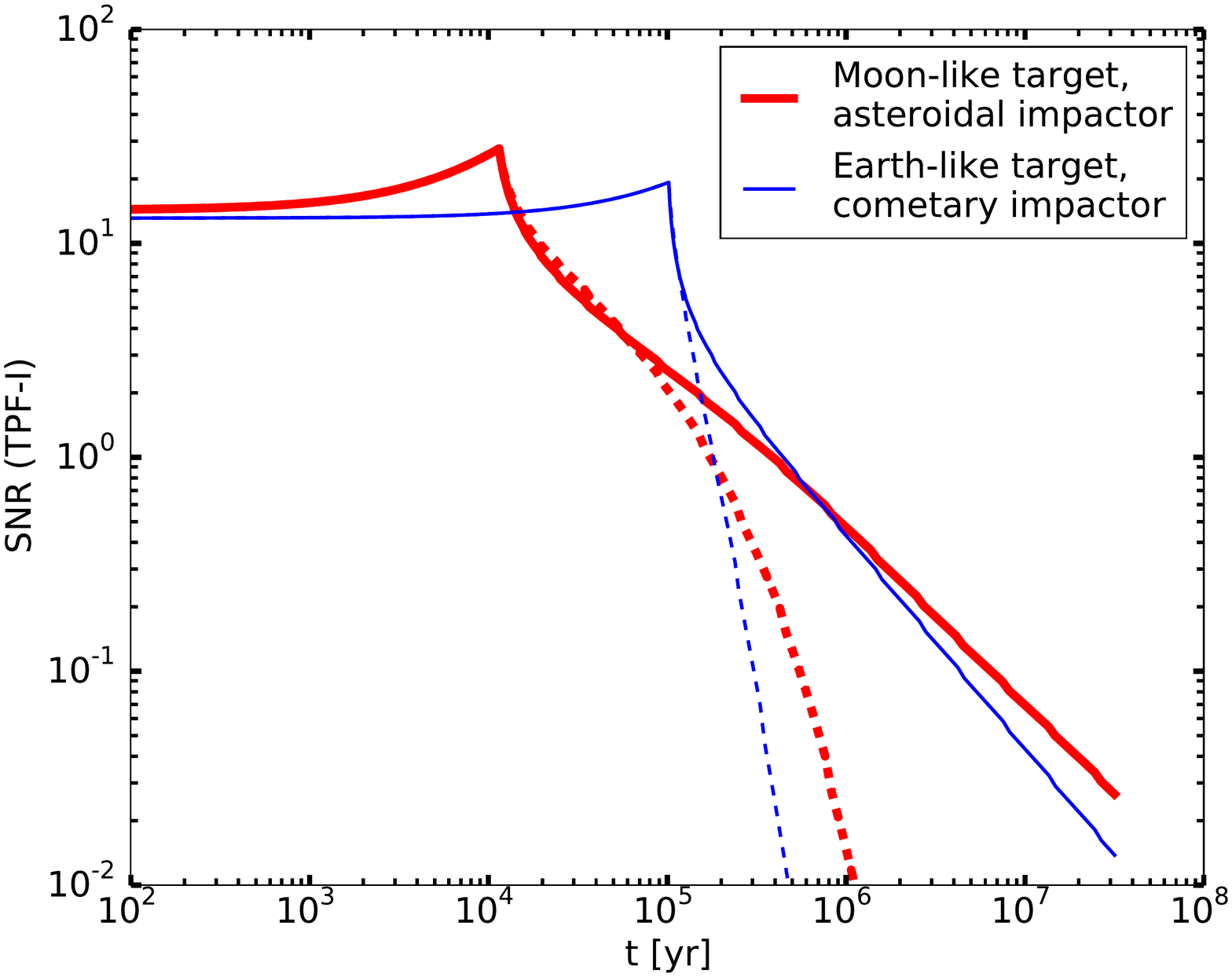}
\caption{The top figure shows the peak flux from silica for two impact scenarios as a function of time after the impact. Dashed lines take dust removal by PR-drag into account, while for the full lines the cascade is always fully populated down to the blowout size. The right axis show the SNR (from equation \ref{SNR_const_sigma}, assuming an exposure time of one hour) for JWST/MIRI, assuming a spectral resolving power $R\sim100$. This is only an approximations (there is no linear correspondence between the peak flux and the SNR), but accurate within a factor of 2. The bottom figure show the corresponding SNR for TPF-I.}
\label{fig:silica_peak_t}
\end{figure}

To calculate a silica emission spectrum, we follow the same approach as for calcite (section \ref{calcite_astrosili}): we use Mie theory with optical constants (with a constant extrapolation into visible wavelengths) for amorphous silica from \citet[][Jena Database of Optical Constants for Cosmic Dust]{Henning_Mutschke_1997}. Emission features are seen at 9, 12 and 20 \,$\mu$m. As a background spectrum, we again consider astrosilicates. Since the aforementioned debris disks contain substantial amounts of silica, we assume, for the sake of argument, that 10\% of the \emph{recondensed} dust consists of silica. Thermal emission features of silica exist in the mid-IR, so we check whether an instrument such as the \textit{James Webb Space Telescope} \citep[JWST,][]{Gardner_etal_2006} with its Mid~Infrared~Instrument\footnote{Sensitivities are taken from the MIRI Pocket Guide v2.2, \url{http://www.stsci.edu/jwst/instruments/miri/docarchive/miri-pocket-guide.pdf}} (MIRI) or TPF-I could detect silica. Figure \ref{fig:SNR_silica} shows the SNR calculated with equation \ref{SNR_const_sigma} for the intermediate case of a cometary body impacting a Mars-like planet. Figure \ref{fig:silica_peak_t} shows how the peak silica flux and the SNR for the aforementioned instruments changes with time after the impact for the extreme case of Earth- and Moon-like targets. These figures suggest that the detection of silica is possible for all the impact scenarios considered in this work when using TPF-I, and potentially even with JWST when considering impacts onto smaller exoplanets and longer exposure times. This is not surprising given that the silica peak flux is of the order of hundreds of microjanskys, while the (background-limited) sensitivity of e.g.\ JWTS~MIRI (at low spectral resolution) is tens of microjansky, and TPF-I would be even more sensitive.

\subsubsection{Example: reflected light from microorganisms}\label{subsubsec:microorganism_detection}
The detection of biological matter in the escaping impact ejecta would be the most direct evidence that the impacted planet hosts a biosphere. Here we consider whether ejected microorganisms could be detected in reflected star light. From laboratory work, we know that reflectance spectra of microorganisms show near-IR absorption features due to water of hydration and amide bonds within proteins \citep{Dalton_etal_2003,Hegde_etal_2015}. An advantage of detecting this features in the debris belt rather than on the planet itself is that the observed spectrum is not affected by the planet's atmosphere \citep{Schwieterman_etal_2015}.

Reflectance spectra of the debris belt could be obtained with instruments such as TPF-C, constructed to directly image Earth-like exoplanets by efficiently blocking direct star light. To estimate the telescope exposure time needed to detect a microbial absorption feature, we use a simple approach \citep{Arnold_etal_2009} and assume that the required SNR\footnote{Note that this is the SNR for the complete signal; it is not the same SNR as in the previous sections.} of the observation (per spectral resolution element) to detect a given absorption feature is
\begin{equation}\label{required_SNR}
\mathrm{SNR}_\mathrm{req}=\frac{1}{\delta\cdot \varsigma}\cdot \chi
\end{equation}
where $\delta$ is the relative depth of the absorption feature, $\varsigma$ is the fraction of the total reflected light that was reflected by microbes and $\chi$ is the desired significance level (put equal to 5 for this calculation). We now assume that the SNR scales (for a fixed observer distance of 10\,pc) as follows:
\begin{equation}\label{SNR_scaling}
\mathrm{SNR}\propto\frac{F\sqrt{t_\mathrm{exp}}}{N_\mathrm{res}\sqrt{R}}
\end{equation}
where $t_\mathrm{exp}$ is the exposure time, $F$ the total flux from the target (in our case the light reflected by both the spalled and the recondensed population), $R$ the resolving power and $N_\mathrm{res}$ the number of spatial resolution elements\footnote{This parameter is used to transform the SNR for observations of exoplanets, where $N_\mathrm{res}=1$, to observations of dust belts with $N_\mathrm{res}>1$.} occupied by the debris belt. We take into account that structures closer to the star than 4$\lambda$/D (with $D$ the diameter of the telescope) are blocked by the coronagraph. As reference values, we follow \citet{Arnold_etal_2009} and use the exposure time estimates by \citet{Guyon_etal_2006} to directly image an Earth-twin with $\mathrm{SNR}=7$ using a coronagraph and a 100\,nm wide band centred at 550\,nm. We now compute the exposure time (using equations \ref{required_SNR} and \ref{SNR_scaling}) necessary to detect the water of hydration feature at 1.5\,$\mu$m \citep{Dalton_etal_2003,Hegde_etal_2015}, a wavelength that might be in reach of a future TPF-C-like telescope. The flux $F$ is scaled by both the different cross-section of the debris belt compared to the Earth-twin and the reduced stellar emission (for an identical bandwidth) at 1.5\,$\mu$m compared to visible wavelengths. The resolving power $R$ is chosen such that the absorption feature is resolved. Note that at $1.5$\,$\mu$m, at least $\sim$50\% of the belt is not observable because of masking by the coronagraph.

We measure a typical depth of the 1.5\,$\mu$m absorption feature of 0.52 from a mean reflectance spectrum computed from the microbial reflectance spectra library by \citet{Hegde_etal_2015}. We also need to put a number on the fraction of microbes in the total (spalled and recondensed) ejecta. First, we assume that only ejecta experiencing shock pressure below 10\,GPa can contain detectable microorganisms, and we only include grains larger than 10\,$\mu$m (i.e.\ significantly larger than a single microbial cell). We then need to estimate the fraction (by volume) of microbes among these spalled, low-pressure ejecta. Again, Earth is our only viable example. For the sake of argument, let us take as a reference value the density of prokaryotic cells in unconsolidated subsurface sediments in the upper 10\,m, which was determined by \citet{Whitman_etal_1998} to be $2.2\times10^8$\,cells/cm$^3$. This corresponds to a microbe fraction (by volume) of $\sim$$2\times10^{-4}$.

Putting all these numbers into equations \ref{required_SNR} and \ref{SNR_scaling}, we conclude that the detection of the 1.5\,$\mu$m water of hydration absorption feature is impossible to achieve within reasonable exposure time, even when considering a 12\,m telescope that could detect an Earth-twin in just two minutes. The major factor leading to the huge exposure times is the relative rareness of microbes among the ejecta (the required SNR is inversely proportional to the volume fraction of microbes). Since we do not know whether Earth is a representative example of a life-harbouring planet, we can imagine worlds that offer more favourable conditions to life than Earth does. However, it has been argue that such ``superhabitable'' worlds would preferentially be more massive than Earth \citep{Heller_Armstrong_2014}, resulting in less escaping material during an impact event. Moreover, calculated exposure times remain long (a month or longer) even when making the extreme assumption that 50\% of the low-pressure ejecta ($<$10\,GPa) are microbial and that in addition the spatial resolution elements are added up (i.e. $N_\mathrm{res}$ is replaced by $\sqrt{N_\mathrm{res}}$ in equation \ref{SNR_scaling}), and otherwise favourable conditions (Moon-sized exoplanet, ignoring PR-drag), regardless at which time after the impact the dust is observed \footnote{The required SNR decreases as the spalled population becomes more dominant, but so does the total flux.}. Note also that our assumption about the SRN scaling linearly with $F$ is optimistic. Because the contrast of the dust is in general higher than the reference contrast of the Earth-twin, the linear dependence increases the SNR more than if $SNR\propto\sqrt{F}$, which would be the case if the shot noise of $F$ contributes significantly to the total noise. Besides unreasonably long exposure times, additional difficulties exist. For example, the water of hydration can be removed by the effect of space vacuum \citep{Dose_2003}.

\subsubsection{Confusion of exoplanetary dust with exozodiacal dust}
Impact events imply a reservoir of impactors that generate dust themselves. In the solar system, this is manifested by the presence of the zodiacal dust in the terrestrial region, which is thought to originate primarily from comets with a smaller contribution from mutually colliding asteroids and interstellar dust \citep{Nesvorny_etal_2010,Rowan-Robinson_May_2013}. Therefore, there is a risk that the signal from an impact event is obscured by exozodiacal dust. It could be imagined that an LHB-like event would accumulate large quantities of impact-generated dust. However, in the solar system, the amount of zodiacal dust was significantly higher as well during the LHB due to the increased influx of cometary bodies \citep{Nesvorny_etal_2010}. We note that exozodiacal dust can also be delivered to the terrestrial region from cool Kuiper belt analogues by PR-drag \citep{Kennedy_Piette_2015}. Exozodiacal dust is a considerable source of noise for observations aimed at direct imaging and spectroscopy of Earth-like exoplanets in the habitable zone. Therefore, the characterisation of the exozodiacal dust luminosity function is of prime interest for the preparation of future missions targeted to the direct detection of Earth twins. Such efforts will be carried out over the next few years and give a much more detailed picture of the prevalence and luminosity distribution of exozodiacal dust. For example, \citet{Weinberger_etal_2015} describe a future survey of $\sim$50 stars with the LBTI searching for exozodiacal dust \citep[see also][]{Kennedy_etal_2015}.

Is there a possibility to distinguish between exozodiacal dust\footnote{We refer to exozodiacal dust as "classical" dust produced from asteroids or comets, but it could be argued that dust from impacts is just another form of exozodiacal dust.} from asteroids or comets and dust from planetary impacts? Certainly, a difference would be expected in dust composition, which can be studied by infrared spectroscopy \citep[e.g.][]{deVries_etal_2012} or reflection spectroscopy \citep[e.g.][]{Debes_etal_2008,Kohler_etal_2008}. Exozodiacal dust is expected to have a cometary or asteroidal composition, while dust from a planetary surface might resemble terrestrial crust (or mantle) material and contain impact-generated silica, which, as seen in section \ref{sec:silica_emission}, have good prospects to be detectable with future instruments. Studies trying to link dust composition to its origin have already been carried out \citep[e.g.][]{Lisse_etal_2012}. \citet{Morlok_etal_2014} presented absorption spectra of materials representative of the Earth crust (e.g.\ granite, basalt) and mantle (e.g.\ dunite) as well as martian meteorites and compared these laboratory data to infrared spectra of debris disks. So, by looking at the dust composition, one should in principle be able to distinguish between asteroidal/cometary dust and exoplanetary dust.

It might also be possible to distinguish between exozodiacal dust and impact generated dust by considering their spatial distribution, but this is beyond the scope of this paper. We nevertheless note that models of the dust density distribution in the zodiacal cloud are quite complex, consisting of several distinct components \citep[e.g.][]{Kelsall_etal_1998,Krick_etal_2012,Rowan-Robinson_May_2013}. The spatial distribution of debris created in a giant impact was studied by \citet{Jackson_Wyatt_2012}.

\subsection{Rate of large-scale impact events}
The rate of impacts violent enough to accelerate target material to escape velocity is an important parameter to discuss. As a first step, consider the impact rate on Earth today. Events similar or larger than the K-T impact (impactor diameter $\sim$10\,km) are expected to occur, on average, every $\sim$100\,Myr \citep[e.g.][]{Harris_2008}. Several tens of such events should therefore have occurred over the lifetime of Earth. Estimates for impactors with a diameter $\geq$20\,km (as considered in section \ref{subsec:ejected_material}) indicate impact intervals of $\sim$500\,Myr \citep{Harris_2008}. This can be compared to the typical lifetime of the post-ejection debris, characterised (if PR-drag is ignored) by $t_\mathrm{max}$ in tables \ref{tab:spalled} and \ref{tab:recondensed}. Assuming that similar impact rates apply to exoplanetary systems, these numbers suggest that only about one in a million systems\footnote{Strictly speaking, these estimates apply only to exoplanetary systems with a solar type host star and a rocky planet at 1\,AU distance, since e.g.\ the collisional evolution depends on the mass and luminosity of the central star.} should show dust from a recondensed population at its maximum fractional luminosity. On the other hand, a few percent of the systems could show dust from a spalled population at its maximum fractional luminosity.

How has the impact rate varied during Earth's history? It is thought that the impact rate was constant from approximately 3.5\,Gyr ago until today \citep[][figure 3]{Valley_etal_2002}. However, in the first billion years after Earth's formation, the impact rate was significantly higher, although different models exist for its exact evolution \citep{Valley_etal_2002}. One possible scenario is a Late Heavy Bombardment (LHB), a spike in the impact rate in the inner solar system 3.8\,Gyr ago, evidence for which is found from e.g.\ Apollo lunar samples. Such a late period of intense bombardment after a relatively quiet phase can be explained by the destabilisation of the orbits of a large number of planetesimals in the outer solar system, caused by the migration of the gas giants \citep{Gomes_etal_2005}. An intense bombardment could increase the chances of detecting debris ejected from a planetary surface, although the level of exozodiacal dust is also expected to be significantly higher during such an event \citep{Nesvorny_etal_2010}. However, was life already present at these early stages? Although the end of the LHB on Earth roughly coincides with the earliest evidence for life, it is indeed possible that life already emerged before the LHB, in the Hadean, and was present throughout the whole period of intense bombardment \citep{Abramov_Mojzsis_2009}. In addition, LHB-like events might occur up to several Gyr after the formation of the planetary system, i.e.\ once life is already well-established on the host planet. This is because the onset of the instability causing the bombardment depends on the initial separation between the outer migrating planet and the planetesimal belt, where larger separation leads to a later LHB \citep{Gomes_etal_2005}. As an example, it has been suggested that dust seen around the 1\,Gyr old main-sequence star $\eta$~Corvi originates from an LHB-like event \citep{Wyatt_etal_2007,Lisse_etal_2012}. However, a caveat to a pre-LHB origin of life is the idea that a high impact rate might be a prerequisite for life to arise in the first place by delivering water and organics \citep[e.g.][]{Pierazzo_Chyba_1999,Court_Sephton_2009} or by inducing hydrothermal vents \citep[e.g.][]{Abramov_Kring_2007}. The latter have been proposed as potential sites for the origin of life \citep[e.g.][]{Baross_Hoffman_1985}. As to the occurrence of LHB-like events in exoplanetary systems, \citet{Booth_etal_2009} estimated from observations that less than 12\% of the Sun-like stars undergo an LHB, i.e.\ extrasolar LHB events are rare.

From ages of impact craters on Earth, there exists evidence that impacts occur in ``bombardment episodes'', possibly due to the breakup of giant comets and subsequent impact of the debris \citep{Napier_2015}. If this is indeed the case, a similar mechanism in an exoplanetary system could help to detect dust from impact events by accumulating the dust of a bombardment episode.

\section{Summary and conclusions}\label{Sec:Conclusions}
We have calculated the mass escaping from an exoplanet during an impact event for a limited number of impact scenarios (exoplanet sizes, impactor types) using a simple model. We have also determined the fraction of escaping ejecta that is not considerably shock damaged and remains in the solid state. We then computed the collisional evolution of the debris with a simplified analytical model based on timescales of collisions (production of new, smaller grains) and removal by PR-drag and radiation pressure. For the relatively small dust masses considered here, PR-drag is an important removal process, in contrast to most known debris disks. We consider two dust populations that we assume to evolve independently of each other: the spalled population that consists of ejecta that remained in the solid state during the impact event, and the recondensed population that formed from melt droplets or vapour. The recondensed population is initially much brighter since it has a larger mass and smaller grain sizes than the spalled population. However, it is also removed faster, such that the spalled population becomes dominant typically after a million years. The fractional luminosity of the impact generated dust is roughly comparable to the fractional luminosity of the zodiacal dust. Such a fractional luminosity is potentially in the reach of the presently available LBTI. Future instruments such as TPF-like telescopes will be able to detect the presence of dust both in thermal emission and scattered light.

By studying the composition of the dust, one would gain information on the impacted exoplanet, its geology or the presence of a biosphere. The escaping masses we derive can be used to assess the detectability of (bio)signatures present in the subsurface of an exoplanet. As examples, we considered three different substances. Calcite would likely indicate the presence of liquid water at some point in the history of the exoplanet, provided the dust indeed originated from a planetary body. Our estimates show that a detection of calcite would require advanced far-IR instruments not yet available. Glassy silica on the other hand would potentially be detectable with JWST, and should be within the reach of a TPF-like telescope. Since glassy silica is linked to violent impacts or collisions, the detection of substantial amounts would suggest that the observed dust indeed originated from an impact event involving a planetary body or large planetesimals. Finally, we considered the direct detection of ejected biological matter (microorganisms) in reflected light. While microorganisms have absorption features due to water of hydration, their detection within the ejected debris seems unfeasible. Indeed, a large fraction of any microorganisms would be destroyed during the impact. In addition, taking Earth as a reference, the density of microorganisms is too low to allow detection.

Our calculations show that ejected dust masses are relatively small. We note however that the dust cross-section is in general larger than the planetary cross-section. Also, studying ejected dust could be an interesting complement to atmospheric studies. Looking at dust from impact events seems to be the only possibility to study the subsurface composition of exoplanets directly, apart from exoplanets evaporating very close to their host star \citep[e.g.][]{vanLieshou_etal_2014}. Dust from impact events is a valuable piece of information with which to achieve a complete characterisation of an exoplanet (including its geology), which is important when assessing its habitability or potential signatures of a biosphere. However, the detailed study of dust from impact events on terrestrial planets has to wait for future instruments with high sensitivity and the ability to observe material in the terrestrial region of a star.

\section*{Acknowledgements}
The authors would like to thank Natalia Artemieva, Axel Brandenburg, Gerda Horneck, Markus Janson and Andreas Morlok for valuable discussions. This research has made use of NASA's Astrophysics Data System. Figures were made using the \texttt{Matplotlib} library \citep{Hunter_2007}.

\section*{Author Disclosure Statement}
No competing financial interests exist.

\def\bibfont{\small}
\bibliography{bibliography}
\bibliographystyle{aa}

\end{document}